\documentclass[a4paper]{article}  

\usepackage{float}
\usepackage[
  a4paper,top=3cm,bottom=2cm,left=3cm,right=3cm,marginparwidth=1.75cm]{geometry}
\usepackage{amsmath}
\usepackage{calrsfs}
\usepackage{amsthm,bm,mathtools,etoolbox,mathrsfs,amsfonts,bbm}
\usepackage{hyperref}
\hypersetup{colorlinks,linkcolor={blue},citecolor={blue},urlcolor={blue}}    
\usepackage[nameinlink,capitalise]{cleveref}
\usepackage[utf8]{inputenc} 
\usepackage[T1]{fontenc}    
\usepackage{url}            
\usepackage{booktabs}       
\usepackage{natbib}
\usepackage{placeins}
\usepackage{adjustbox}
\usepackage{booktabs}
\usepackage{algorithm}
\usepackage{algpseudocode}
\usepackage{multicol} 
\usepackage{makecell, multirow, tabularx} 
\usepackage{rotating}
\usepackage{float}
\usepackage{ulem}
\usepackage{graphicx}
\usepackage{tabularx}
\usepackage{authblk}

\newtheorem{Theorem}{Theorem}[section]

\newtheorem{Lemma}{Lemma}[section]
\newtheorem{Remark}{Remark}[section]
\newtheorem*{Remarknn}{Remark}
\newtheorem{Corollary}{Corollary}[section]
  
\title{Multi-step ahead prediction intervals for non-parametric 
autoregressions via bootstrap: consistency, debiasing and pertinence }

\author[1]{Dimitris N. Politis}
\author[2]{Kejin Wu}

\affil[1]{Department of Mathematics and Halicio\u{g}lu Data Science Institute, University of California, San Diego}
\affil[2]{Department of Mathematics, University of California, San Diego }

\date{}

\begin{document}

\maketitle

\begin{abstract}
To address the difficult problem  of multi-step ahead prediction
of non-parametric autoregressions, we consider a  forward bootstrap approach.  
Employing a local constant estimator,  we can analyze a general type of 
non-parametric time series model, and show that the proposed point predictions are consistent with the true optimal predictor. We construct a quantile prediction interval that is asymptotically valid. Moreover, using a debiasing technique, we can asymptotically approximate the distribution of multi-step ahead non-parametric estimation by bootstrap. As a result, we can build bootstrap prediction intervals that are pertinent, i.e., 
  can capture the model estimation variability, thus improving 
upon the standard quantile prediction intervals.
 Simulation studies are given to illustrate the performance of our point predictions and pertinent prediction intervals for finite samples.
\end{abstract}

\section{Introduction}\label{Sec:Intro}
To model the asymmetry in financial returns, volatility of stock markets,  switching regimes, etc.,   non-linear time series models have attracted 
attention since the 1980s. Compared to linear time series models, non-linear models possess more capabilities to depict the underlying data-generating mechanism; % especially when prior information indicates that the non-linear model is natural to describe data or there are non-linearities existing in data;
 see the review of \cite{politis2009financial} for example. However, unlike linear models where the one-step ahead predictor can be iterated,
multi-step ahead prediction of non-linear models is cumbersome, since the innovation influences the forecasting value severely.

 In this paper, by combining the forward bootstrap 
of \cite{politis2015model} with non-parametric estimation, we develop 
multi-step ahead (conditional) predictive inference for the 
general model:
\begin{equation}\label{Eq:1}
    X_t = m(X_{t-1},\ldots, X_{t-p}) + \sigma(X_{t-1},\ldots, X_{t-q})\epsilon_t;
\end{equation}
 here, the $\epsilon_t$ are assumed to be independent, identically distributed (i.i.d.) with mean 0 and   variance 1, and the $m(\cdot)$ and $\sigma(\cdot)$ are some functions that satisfy some smoothness conditions. 
We will also assume that the time series
  satisfying \cref{Eq:1} is geometrically ergodic  and causal, i.e., 
that for any $t$, $\epsilon_t$ is independent of $\{X_s, s<t\}$. 

In \cref{Eq:1}, we have the trend/regression function $m(\cdot)$ depending on the last $p$ data points, while the  standard deviation/volatility function $\sigma(\cdot)$ depends on the last $q$ data points;
in many situations, $p$ and $q$ are taken to be equal for simplicity. Some special cases deserve mention, e.g., if $\sigma(X_{t-1},\ldots, X_{t-q})$ $\equiv \sigma$ (constant), \cref{Eq:1} yields a non-linear/non-parametric autoregressive model with homoscedastic innovations.
The well-known ARCH/GARCH models are
a special case of  \cref{Eq:1} with $m(X_{t-1},\ldots, X_{t-p})\equiv 0$.

Although the $L_2$ optimal one-step ahead prediction of \cref{Eq:1} is trivial when we know the regression function $m(\cdot)$ or have a consistent estimator of it, the multi-step ahead prediction is not easy to obtain. In addition, it is non-trivial to find the $L_1$ optimal prediction even for the one-step ahead forecasting. In several applied areas, e.g. econometrics, 
climate modeling, water resources management, etc., data might not possess a finite 2nd moment in which case optimizing $L_2$ loss is vacuous; For all such cases---but also of independent interest---prediction that is optimal with respect to $L_1$ loss should receive more attention in practice; see detailed discussions from Ch. 10 of \cite{politis2015model}. Later, we will show our method is compatible with both $L_2$ and $L_1$ optimal multi-step ahead predictions.

The efforts to overcome the difficulty of forecasting non-linear time series could be traced back to the work of \cite{pemberton1987exact}, where a numerical approach was proposed to explore the exact conditional $k$-step ahead $L_2$ optimal prediction of $X_{T+k}$ for a homoscedastic \cref{Eq:1}. However, this method is 
intractable computationally with long-horizon prediction, and
requires knowledge of the distribution of innovations and the regression function which is  not realistic in practice. 

Consequently, practitioners started to investigate some suboptimal methods to
perform multi-step ahead prediction. Generally speaking, these methods   take one of two avenues: (1) Direct prediction or (2) Iterative prediction.
The first idea involved  working with a different (`direct')   model, 
specific to $k$-step ahead prediction, namely:
\begin{equation}
\label{Eq:1d}
    X_t = m_{k}(X_{t-k},\ldots,X_{t-k-p+1}) + \sigma_{k}
(X_{t-k},\ldots,X_{t-k-q+1})\xi_t.
\end{equation}
%\footnote{you had $X_t = m_{d}(X_{t-k},\ldots,X_{t-k-p+1}) + \sigma_{d}
%(X_{t-k},\ldots,X_{t-k-q+1})\xi_t,$ and I removed the +1}
Even though $m_{k}(\cdot)$ and $\sigma_{k}(\cdot)$ are unknown to us, we can  construct non-parametric estimators $\widehat{m}_{k}$ and $\widehat{\sigma}_{k}$, and plug them in \cref{Eq:1d} to perform
$k$-step ahead prediction.
\cite{lee2003new} give a review of this approach. However, as pointed out by \cite{chen2004nonparametric}, a drawback of this approach is that the information of intermediate observations $\{X_{t},\ldots,X_{t-k+1}\}$ is disregarded.
Furthermore, if the $\epsilon_t$ of \cref{Eq:1}  are i.i.d., then
the $\xi_t$ of \cref{Eq:1d} can not be i.i.d.  In other words, a practitioner
must employ the (estimated) dependence structure of the $\xi_t$ of \cref{Eq:1d}
in order to perform the prediction in an optimal fashion

  The second idea is ``iterative prediction'' which employs 
one-step ahead predictors in a sequential way, to perform
a multi-step ahead forecast. For example, consider a 2-step ahead prediction
under model \cref{Eq:1};    first note that the $L_2$ optimal   predictor of 
$X_{T+1}$ is  $\hat X_{T+1}=m(X_{T},\ldots,X_{T+1-p})$.
The $L_2$  optimal predictor of $X_{T+2} = m(X_{T+ 1},X_T,\ldots,X_{T+2-p})$ but 
  since $X_{T+ 1}$ is unknown, it is tempting to plug-in $\hat X_{T+1}$
in its place.  This plug-in idea can be extended to multi-step ahead forecasts
but it does {\it not} lead to the $L_2$ optimal predictor except in the special 
case where the function $m(\cdot)$ is linear, e.g., in the case 
  of a linear auto-regressive (LAR) model.

\begin{Remarknn} \label{Re.Intro}
Since neither of the above two approaches is satisfactory, we propose to approximate the distribution of the future value via  a particular type
of {\it simulation}  when the model
is known or, more generally, by {\it bootstrap}. To describe this approach, we rewrite \cref{Eq:1} as 
$$X_{t} = G(\bm{X}_{t-1},\epsilon_t)$$
 where $\bm{X}_{t-1}$ is a vector which represents $\{X_{t-1},\ldots,X_{t-\text{max}(p,q)}\}$ and $G(\cdot,\cdot)$ is some appropriate function.
  Then, when model and innovation information are known to us, we can create a pseudo value $X^*_{T+k}$. Take a three-step ahead prediction as an example, the pseudo value $X^*_{T+3}$ can be defined as below:
\begin{equation}\label{Eq:3}
    X^*_{T+3} =  G(G(G(\bm{X_T},\epsilon^*_{T+1}),\epsilon^*_{T+2}),\epsilon^*_{T+3});
\end{equation}
here $\{\epsilon^*_i\}_{i=T+1}^{T+3}$ are simulated as i.i.d. from $F_{\epsilon}$. Repeating this process to $M$ pseudo $X^*_{T+3}$, the $L_2$ optimal prediction of $X_{T+3}$ can be estimated by the mean of $\{X^{*(m)}_{T+3}\}_{m=1}^{M}$. 
As already discussed, constructing the $L_1$ optimal predictor may also be required since sometimes $L_2$ loss is not well-defined; in our simulation framework, we can construct the optimal $L_1$ prediction by taking the median of $\{X^{*(m)}_{T+k}\}_{m=1}^{M}$. Moreover, we can even build a prediction interval (PI) to measure the forecasting accuracy based on quantile values of simulated pseudo values. The extension of this algorithm to longer step ahead prediction is illustrated in \cref{Sec:non-parametric_bootstrap}.  
\end{Remarknn}
% \footnote{Actually I see in Section 2 you define the algorithms specific
% to the additive model of eq. (\ref{Eq:1}).
% Maybe define the algorithms for the general
%  eq. (\ref{Eq:3}) in the beginning ot Section 2   and then say that to 
% prove consistency etc we now have to  limit ourselves to
%  the  additive model of eq. (\ref{Eq:1}). 
%    }

Realistically, practitioners would not know $F_{\epsilon}$, $m(\cdot)$ and $\sigma(\cdot)$. In this situation, the first step is to estimate these quantities and
plug them into the above simulation  which then
 turns into  a bootstrap method. In the spirit of this idea, some studies were done by adopting different bootstrap techniques. \cite{thombs1990bootstrap} proposed a \textit{backward bootstrap} trick to predict $AR(p)$ model. The advantage of the backward method is that each bootstrap prediction is naturally conditional on the latest $p$ observations which coincide with the conditional prediction in the real world. However, this method can not handle non-linear time series whose backward representation may not exist. Later, \cite{pascual2004bootstrap} proposed a strategy to generate bootstrap $AR(p)$ series forward. For resolving the conditional prediction issues, they fixed the last $p$ bootstrap values to be the true observations and compute predictions iteratively in the bootstrap world starting from there. They then extended this procedure to forecast the GARCH model in \cite{pascual2006bootstrap}. 

Sharing a similar idea, \cite{pan2016bootstrap} defined the \textit{forward bootstrap} to do prediction, but they proposed a different PI format which empirically has better performance according to the coverage rate (CVR) and the length (LEN), compared to the PI of \cite{pascual2004bootstrap}. Although \cite{pan2016bootstrap} covered the forecasting of a non-linear
and/or non-parametric  time series model, only one-step ahead prediction was considered. 
The case of multi-step ahead prediction of non-linear (but parametric)
time series models was recently addressed in \cite{wu2023bootstrap} 
%broadened the \textit{forward bootstrap} to multi-step ahead predictions of the model \cref{Eq:1} with generally parametric non-linear $m(\cdot)$ and $\sigma(\cdot)$; see the book \cite{KreissPaparoditisbook2023} for more applications of bootstrap on time series.
In the paper at hand, we address the  case of multi-step ahead prediction
of non-parametric time series models as in \cref{Eq:1}.   Beyond discussing optimal $L_1$ and $L_2$ point predictions, we consider two types of PI---Quantile PI (QPI) and Pertinent PI (PPI). 
As already mentioned,  the former can be approximated by taking quantile values of the future value's distribution in the bootstrap world. The PPI requires a more complicated and computationally-heavy procedure to be built as it attempts to capture the variability of parameter estimation. This additional effort results in improved finite-sample coverage as compared to the QPI.

As in most non-parametric estimation problems, the issue of bias becomes important.
 We will show that debiasing on the inherent bias-type terms of non-parametric estimation is necessary to guarantee the pertinence of a PI when multi-step ahead predictions are required. Although QPI and PPI are asymptotically equivalent, the PPI renders better CVR in finite sample cases; see the formal definition of PPI in the work of \cite{politis2015model} and \cite{pan2016bootstrap}. Analogously to the successful construction of PIs  in the work of \cite{politis2013model}, we may  employ predictive—as opposed to fitted—residuals in the bootstrap process to further alleviate the finite-sample undercoverage of bootstrap PIs in practice.

The paper is organized as follows. In \cref{Sec:non-parametric_bootstrap}, forward bootstrap prediction algorithms with local constant estimators will be given. The asymptotic properties of point predictions and PIs will be discussed in \cref{Sec:asymptotic}. Simulations are given in \cref{Sec:simulation} to substantiate  the finite-sample performance of our methods. Conclusions are given in \cref{Sec:conclusion}. All proofs can be found in \hyperref[Appendix:Proof]{Appendix A}. Discussions on the debiasing and pertinence related to building PIs are presented in \hyperref[Appendix:advanQPI]{Appendix B} to \hyperref[Appendix:optbandwidthonestvar]{Appendix D}.

\section{Non-parametric forward bootstrap prediction}\label{Sec:non-parametric_bootstrap}
 As  discussed in Remark \ref{Re.Intro}, we can apply the \textit{simulation} or \textit{bootstrap} technique to approximate the distribution of a future value. In general, this idea works for any geometrically ergodic auto-regressive model no matter in a linear or non-linear format. For example, if we have a known general model $X_{t} = G(\bm{X}_{t-1},\epsilon_t)$ at hand, we can do $k$-step ahead predictions according to the same logic of the three-step ahead prediction example of Remark \ref{Re.Intro}.

To elaborate, we  need to simulate $\{\epsilon^*_i\}_{i=T+1}^{T+k}$ as i.i.d. from $F_{\epsilon}$ and then compute pseudo value $X_{T+k}^*$ iteratively with simulated innovations as below:
\begin{equation}\label{Eq:4}
    X^*_{T+k} =  G(\cdots G(G(G(\bm{X_T},\epsilon^*_{T+1}),\epsilon^*_{T+2}),\epsilon^*_{T+3}),\ldots,\epsilon^*_{T+k})
\end{equation}
Repeating this procedure $M$ times, we can make prediction inference with the empirical distribution of $\{X^{*(m)}_{T+k}\}_{m=1}^{M}$. 
%For a more detailed description, we refer to Section 2.1 of \cite{wu2023bootstrap}.
 Similarly, if the model and innovation distribution are unknown to us, we can do the estimation first to get $\widehat{G}(\cdot,\cdot)$ and $\widehat{F}_{\epsilon}$. Then, the above simulation-based algorithm turns out to be a bootstrap-based algorithm. More specifically, we bootstrap $\{\hat{\epsilon}^*_i\}_{i=T+1}^{T+k}$ from $\widehat{F}_{\epsilon}$ and calculate pseudo value $\widehat{X}_{T+k}^*$ iteratively with $\widehat{G}(\cdot,\cdot)$. The prediction inference can also be conducted with the empirical distribution of $\{\widehat{X}^{*(m)}_{T+k}\}_{m=1}^{M}$.
% We refer to Algorithms 1 and 2 of \cite{wu2023bootstrap} for more details. 

The simulation/bootstrap idea of Remark \ref{Re.Intro} was recently implemented 
by  \cite{wu2023bootstrap} in the case where the model $G$ is either known 
or parametrically specified. In what follows, we will focus on the case of
a   non-parametric model \cref{Eq:1} and will  analyze the asymptotic properties of the point predictor and prediction interval.  For the sake of simplicity, we consider only the case $p=q=1$; the general case  can be handled similarly but the notation is quite more cumbersome. Assume we observe $T+1$ datapoints and we denote them as $\{X_{0},\ldots,X_{T}\}$; our goal is prediction inference of $X_{T+k}$ for some $k\geq1$. If we know $m(\cdot)$, $\sigma(\cdot)$ and $F_{\epsilon}$, we can take a simulation approach to develop prediction inference as we explained in \cref{Sec:Intro}. %; see Section 2 of Wu and Politis (2023) for a more detailed discussion. 
When $m(\cdot)$, $\sigma(\cdot)$ and $F_{\epsilon}$ are
unknown, we start by estimating $m(\cdot)$ and $\sigma(\cdot)$;
we then estimate $F_{\epsilon}$ based on the empirical distribution of 
residuals. Subsequently, we can deploy a bootstrap-based method to approximate the distribution of future values. Several algorithms are given for this purpose in the later context.  

\subsection{Bootstrap algorithm for point prediction and QPI}
For concreteness, we focus on local constant estimators, 
i.e., kernel-smoothed estimators of Nadaraya-Watson type;
other estimators can be applied similarly. The  local constant estimators of $m(\cdot)$ and $\sigma(\cdot)$ are respectively defined as:
\begin{equation}\label{estimator}
    \widetilde{m}_h(x) =  \frac{\sum_{t=1}^{T}K(\frac{x-X_{t-1}}{h})X_{t}}{\sum_{t=1}^{T}K(\frac{x-X_{t-1}}{h})} ~~\text{and}~~ \widetilde{\sigma}_{h}(x) = \frac{\sum_{t=1}^{T}K(\frac{x-X_{t-1}}{h})(X_{t}-\widetilde{m}_h(X_{t-1}))^2}{\sum_{t=1}^{T}K(\frac{x-X_{t-1}}{h})};
\end{equation}
here, $K$ is a non-negative kernel function that 
satisfies some regularity assumptions; see \cref{Sec:asymptotic} for details. 
We use $h$ to represent the bandwidth of kernel functions but $h$ may take a different value for mean and variance estimators. Due to the theoretical and practical issues, we need to truncate the above   local constant estimators as follows:
\begin{equation}\label{truncatedest}
    \widehat{m}_{h}(x) = \begin{cases} 
    -C_{m}&\text{if}~\widetilde{m}_h(x)< -C_m\\
    \widetilde{m}_h(x)&\text{if}~|\widetilde{m}_h(x)|\leq C_m\\
    C_{m}&\text{if}~\widetilde{m}_h(x)> C_m\\
    \end{cases}~;~\widehat{\sigma}(x) = \begin{cases} 
    c_{\sigma}&\text{if}~\widetilde{\sigma}_h(x)< c_\sigma\\
\widetilde{\sigma}_h(x)&\text{if}~c_{\sigma}\leq\widetilde{\sigma}_h(x)\leq C_\sigma\\
    C_{\sigma}&\text{if}~\widetilde{\sigma}_h(x)> C_\sigma\\
    \end{cases};
    \end{equation}
here, $C_m$ and $C_\sigma$ are large enough and $c_{\sigma}$ is small enough. 

Using  $\widehat{m}_h(\cdot)$ and $\widehat{\sigma}_h(\cdot)$
on \cref{Eq:1}, we can  obtain the fitted residuals $\{\hat{\epsilon}_{t}\}_{t=1}^{T}$ which is defined as:
\begin{equation}\label{fittedres}
    \hat{\epsilon}_{t} = \frac{X_t - \widehat{m}_h(X_{t-1})}{\widehat{\sigma}_h(X_{t-1})},~\text{for}~t = 1,\ldots,T.
\end{equation}
Later in \cref{Sec:asymptotic}, we will show that the innovation distribution $F_{\epsilon}$ can be consistently estimated by the centered empirical distribution of $\{\hat{\epsilon}_{t}\}_{t=1}^{T}$, i.e., $\widehat{F}_{\epsilon}$, under some standard assumptions. 
We now have all the ingredients to perform the bootstrap-based \cref{algori1} to yield the point prediction and QPI of $X_{T+k}$.
\begin{algorithm}[htbp]
\caption{Bootstrap prediction of $X_{T+k}$ with fitted residuals} 
%\centering
\label{algori1}
  %\centering
  \begin{tabularx}{\textwidth}{lX}   
    Step 1 & With data $\{X_{0},\ldots,X_{T}\}$, construct the estimators $\widehat{m}_h(x)$ and $\widehat{\sigma}_h(x)$ by formula \cref{truncatedest}. \\
    Step 2 & Compute fitted residuals based on \cref{fittedres}, and
let $\bar{\epsilon}= \frac{1}{T}\sum_{i=1}^{T}\hat{\epsilon}_i$.  
Denote $\widehat{F}_{\epsilon}$  the empirical distribution of the
centered residuals $\hat{\epsilon}_t -\bar{\epsilon}
$ for $ t= 1,\ldots,T$ .\\
    Step 3 &   Generate $\{\hat{\epsilon}^*_{i}\}_{i=T+1}^{T+k}$   i.i.d.~from $\widehat{F}_{\epsilon}$. Then, construct bootstrap 
 pseudo-values $X^*_{T+1}$, $\cdots, X^*_{T+k}$   iteratively, i.e.,
    \begin{equation}
            X^*_{T+i} = \widehat{m}_h(X^*_{T+i-1}) + \widehat{\sigma}_h(X^*_{T+i-1})\hat{\epsilon}^*_{T+i},~\text{for}~i = 1,\ldots,k.
    \end{equation}
  For example,   $X^*_{T+1} = \widehat{m}_h(X^*_{T }) + \widehat{\sigma}_h(X^*_{T })\hat{\epsilon}^*_{T+1},$ and 
$X^*_{T+2} = \widehat{m}_h(\widehat{m}_h(X_T)+\sigma_h(X_T)\hat{\epsilon}^*_{T+1}) + \widehat{\sigma}_h(\widehat{m}(X_T)+\widehat{\sigma}_h(X_T)\hat{\epsilon}^*_{T+1})\hat{\epsilon}^*_{T+2}$. \\
    Step 4 & Repeating Step 3 $M$ times, we obtain
pseudo-value replicates of $  X^*_{T+k}$ that we denote 
$\{X_{T+k}^{(1)},\ldots,X_{T+k}^{(M)}\}$. Then, $L_2$ and $L_1$ optimal predictors can be approximated by $\frac{1}{M}\sum_{i = 1}^{M}X_{T+k}^{(i)}$ and $\text{Median of }\{X_{T+k}^{(1)},\ldots,X_{T+k}^{(M)}\}$, respectively. Furthermore, a $(1-\alpha)100$\% QPI can be built 
as $(L,U)$ where $L$ and $U$
denote the $ \alpha/2$ and $ 1-\alpha/2$
sample quantiles of $M$ values $\{X_{T+k}^{(1)},\ldots,X_{T+k}^{(M)}\}$.
  \end{tabularx}
\end{algorithm}

\begin{Remarknn}
 To construct the QPI of \cref{algori1}, we may employ the optimal 
bandwidth rate, i.e., $h = O(T^{-1/5})$. However, in practice with small sample size, the QPI has a better empirical CVR for multi-step ahead predictions by adopting an under-smoothing bandwidth; see \hyperref[Appendix:advanQPI]{Appendix B} for related discussions and see \cref{Sec:simulation} for simulation comparisons between applying optimal and under-smoothing bandwidths on QPI.
\end{Remarknn}

In the next section, we will show the conditional asymptotic 
consistency of our optimal point predictions and QPI. In particular, we verify that our point predictions converge to oracle optimal point predictors in probability --conditional on $X_T$.
%\footnote{NO--what we want is our bootstrap point predictor to 
%be asymptotically close to the oracle optimal point predictor }
In addition, we look for an asymptotically valid PI with $(1-\alpha)100\%$ CVR to measure the prediction accuracy conditional on the latest observed data, which is defined as:
\begin{equation}
    \mathbb{P}(L \leq X_{T+k} \leq U) \to 1-\alpha,~\text{as}~T\overset{}{\to}\infty,
\end{equation}
where $L$ and $U$ are lower and higher PI bounds, respectively. Although not
explicitly denoted, the probability $\mathbb{P}$ should be understood as conditional probability given $X_T$. Later, based on a sequence of sets that contains the observed sample with a probability tending to 1, we will show how to build a prediction interval that is asymptotically valid by the bootstrap technique even for model information being unknown. 

% \vspace{20pt}
% {\color{cyan} Later, we will show that this prediction interval can be obtained with $\textit{high probability}$ undertaking the bootstrap technique even for model information being unknown.} What I want to say is that we can find a prediction interval such that
% \begin{equation*}
%    \mathbb{P}\left(\mathbb{P}(\widehat{L} \leq X_{T+k} \leq \widehat{U}) \to 1-\alpha\right) \to 1,~\text{as}~T\overset{}{\to}\infty;
% \end{equation*}
% $\widehat{L}$ and $\widehat{U}$ are determined by the quantile values of the distribution of $X^*_{T+K}$ which can be approximated by bootstrap. The $\textit{high probability}$ is due to the result:
% \begin{equation*}
%     \sup_{|x|\leq c_{T}}\left| F_{X^*_{T+k}|X_T,\ldots,X_0}(x) -  F_{X_{T+k}|X_T}(x)\right| \overset{p}{\to} 0,~\text{for}~k \geq 1,\end{equation*}
% so $F_{X_{T+k}^*}(u) - F_{X_{T+k}^*}(l)$ converges to $F_{X_{T+k}}(u) - F_{X_{T+k}}(l)$ in probability for suitable $u$ and $l$. When $T$ is large enough, we can let $U = F_{X_{T+k}}(u)$ and $L = F_{X_{T+k}}(l)$ such that $\mathbb{P}(L \leq X_{T+k} \leq U) = 1 - \alpha$. And if we denote $\widehat{U} = F_{X^*_{T+k}}(u)$ and $\widehat{L} = F_{X^*_{T+k}}(l)$, we have $\mathbb{P}(\widehat{L} \leq X_{T+k} \leq \widehat{U})$ converges to $ 1 - \alpha$ in probability, i.e., the coverage probability converges to a specific level in probability (or with high probability).
% \vspace{20pt}

Although asymptotically correct, in finite samples the QPI typically suffers 
from undercoverage; see the discussion in \cite{politis2015model} and \cite{pan2016bootstrap}. To improve the CVR in practice, we consider taking the predictive residuals to boost the bootstrap process. To derive such predictive residuals, 
%which are denoted by $\hat{\epsilon}_{t}^{p}$ hereafter, 
we need to estimate the model based on the delete-$X_t$ dataset, i.e., the available data for the scatter plot of $X_{i}$ vs. $\{X_{i-1}\}$ for $i = 1,\ldots,t-1,t+1,\ldots, T$, i.e., excludes the single point at $i = t$. More specifically, we define the   delete-$X_{t}$ local constant estimator as:
\begin{equation}\label{estimator-pre-res}
    \widetilde{m}^{t}_h(x) =  \frac{\sum_{i=1, i\neq t}^{T}K(\frac{|x-X_{i-1}|}{h})X_{i}}{\sum_{i=1, i\neq t}^{T}K(\frac{|x-X_{i-1}|}{h})} ~~\text{and}~~ \widetilde{\sigma}^{t}_h(x) = \frac{\sum_{i=1, i\neq t}^{T}K(\frac{|x-X_{i-1}|}{h})(X_{i}-\widetilde{m}^t_h(X_{i-1}))^2}{\sum_{i=1, i\neq t}^{T}K(\frac{|x-X_{i-1}|}{h})}.
\end{equation}
Similarly, the truncated delete-$X_{t}$ local estimator $\widehat{m}^{t}(x)$ and $\widehat{\sigma}^{t}_h(x)$ can be defined according to \cref{truncatedest}. We
now construct the so-called predictive residuals as:
\begin{equation}\label{preres}
    \hat{\epsilon}^p_{t} = \frac{X_t - \widehat{m}^{t}_h(X_{t-1})}{\widehat{\sigma}^{t}_h(X_{t-1})},~\text{for}~t=1,\ldots,T.
\end{equation}
%we can get $\{\hat{\epsilon}^p_t\}_{t=1}^{T}$. 
The $k$-step ahead prediction of $X_{T+k}$ with predictive residuals is depicted in \cref{algori2}. Although \cref{algori1,algori2} are asymptotically equivalent, \cref{algori2} gives a QPI with better CVR for finite samples; see the simulation comparisons of these two approaches in \cref{Sec:simulation}. 
\begin{algorithm}[htbp]
\caption{Bootstrap prediction of $X_{T+k}$ with predictive residuals} 
%\centering
\label{algori2}
  %\centering
  \begin{tabularx}{\textwidth}{lX}   
    Step 1 & Same with Step 1 of \cref{algori1}.\\
    Step 2 & Compute predictive residuals based on \cref{preres}. 
denote $\widehat{F}^p_{\epsilon}$ the empirical distribution of the
centered predictive  residuals $\hat{\epsilon}^p_t - \frac{1}{T}\sum_{i=1}^{T}\hat{\epsilon}^p_i, t= 1,\ldots,T$.\\
    Steps 3-4 &  Replace $\widehat{F}_{\epsilon}$ by $\widehat{F}^{p}_{\epsilon}$ in \cref{algori1}. All the rest are  the same. 
  \end{tabularx}
\end{algorithm}

\FloatBarrier
\subsection{Bootstrap algorithm for PPI}
To improve the CVR of a PI, we can try to take the variability of the model estimation into account when we build the PI, i.e., we need to mimic the estimation process in the bootstrap world. Employing this idea results in a Pertinent PI (PPI) as discussed in \cref{Sec:Intro}; see also \cite{wang2021model}. 

\cref{algori3} outlines the procedure to build a PPI. Although this algorithm is more computationally heavy, the advantage is that PPI gives better CVR compared to QPI in practice, i.e., with finite samples; see the examples in \cref{Sec:simulation}.  

\begin{algorithm}[H]
\caption{Bootstrap PPI of $X_{T+k}$ with fitted residuals} 
%\centering
\label{algori3}
  %\centering
  \begin{tabularx}{\textwidth}{lX}   
    Step 1 & With data $\{X_{0},\ldots,X_{T}\}$, construct the estimators $\widehat{m}_h(x)$ and $\widehat{\sigma}_h(x)$ by formula \cref{truncatedest}. Furthermore, compute fitted residuals based on \cref{fittedres}. Denote the empirical distribution of centered residuals  $\hat{\epsilon}_t - \frac{1}{T}\sum_{i=1}^{T}\hat{\epsilon}_i, t= 1,\ldots,T$ by  $\widehat{F}_{\epsilon}$.\\
    Step 2 & Construct the   $L_1$ or $L_2$  prediction $\widehat{X}_{T+k}$ using \cref{algori1}.\\
    Step 3 & (a) Resample (with replacement) the residuals from $\widehat{F}_{\epsilon}$ to create pseudo-errors $\{\hat{\epsilon}^{*}_{i}\}_{i=1}^{T}$ and $\{\hat{\epsilon}^{*}_{i}\}_{i=T+1}^{T+k}$.\\
    &(b) Let $X_{0}^*={X}_{I}$ where $I$ is generated as a discrete random variable uniformly on the values $0,\ldots, T$. Then, create 
 bootstrap pseudo-data $\{X_t^*\}_{t = 1}^{T}$ in a recursive manner
from the formula  
 \begin{equation} \label{eq.11}
     X^*_{i} = \widehat{m}_g(X^*_{i-1}) + \widehat{\sigma}_g(X^*_{i-1})\hat{\epsilon}^*_{i},~\text{for}~i = 1,\ldots,T.
 \end{equation}
 \\
 & (c) Based on the bootstrap data $\{X^{*}_{t}\}_{t=0}^{T}$, re-estimate the regression and variance functions according to \cref{truncatedest} and get $\widehat{m}_{h}^*(x)$ and $\widehat{\sigma}_{h}^*(x)$; we use the same bandwidth $h$ as the original estimator $\widehat{m}_h(x)$.\\
 & (d) Guided by the idea of forward bootstrap, re-define the latest value $X_{T}^*$ to match the original, i.e., re-define $X_T^* = X_T$.\\
& (e) With estimators $\widehat{m}_{g}(x)$ and $\widehat{\sigma}_{g}(x)$, the bootstrap data $\{X_t^*\}_{t = 0}^{T}$, and the pseudo-errors $\{\hat{\epsilon}_{t}^*\}_{t = T+1}^{T+k}$, use \cref{eq.11} 
  to generate recursively the future bootstrap data $X_{T+1}^*,\ldots,X_{T+k}^{*}$.\\
  & (f) With bootstrap data $\{X^*_{t}\}_{t=0}^{T}$ and estimators $\widehat{m}_{h}^*(x)$ and $\widehat{\sigma}_{h}^*(x)$, utilize \cref{algori1} to compute the optimal bootstrap prediction which is denoted by $\widehat{X}^*_{T+h}$; to generate bootstrap
 innovations, we still use $\widehat{F}_{\epsilon}$.\\
& (g) Determine the bootstrap predictive root: $X_{T+k}^*- \widehat{X}^*_{T+k}$. \\ 
Step 4 & Repeat Step 3 $B$ times; 
the $B$ bootstrap root replicates are collected in the form of an empirical distribution whose $\beta$-quantile is denoted $q(\beta)$.
The $(1-\alpha)100\%$ equal-tailed prediction interval for $X_{T+k}$ centered at $\widehat{X} _{T+k}$ is then estimated by
 $[\widehat{X}_{T+k}+q(\alpha/2), \widehat{X} _{T+k}+q(1-\alpha/2)].$
  \end{tabularx}
\end{algorithm}
\begin{Remark}[Bandwidth choices]\label{Remark:suboprate}
    In Step 3 (b) of \cref{algori3}, we may use an
optimal bandwidth $h$ and an  over-smoothing bandwidth $g$ to generate bootstrap time series so that we can capture the asymptotically non-random bias-type term of non-parametric estimation by the forward bootstrap; see the application in \cite{franke2002bootstrap}. We can also apply
an  under-smoothing bandwidth $h$ (and then use $g=h$) to render the
 bias term negligible. It turns out that
both approaches % taking over-smoothing or under-smoothing bandwidth works
work well for one-step ahead prediction, although applying the over-smoothing bandwidth may be slightly better. 
However, taking under-smoothing bandwidth(s) is notably better for multi-step ahead prediction. The reason for this is that the bias term can not be captured appropriately for multi-step ahead estimation with over-smoothing bandwidth. On the other hand, with under-smoothing bandwidth  the bias term is negligible; see \cref{Subsec:PPI} for more discussion---also see \cite{politisstudentization} for related discussion. The simulation studies in \hyperref[Appendix:diffoverunderband]{Appendix C}  explore the differences between these two bandwidth strategies.  
\end{Remark}

As \cref{algori2} was a version of \cref{algori1} using  
predictive (as opposed to fitted) residuals, we now propose \cref{algori4} 
that constructs a PPI with predictive residuals. 

\begin{algorithm}[htbp]
\caption{Bootstrap PPI of $X_{T+k}$ with predictive residuals} 
%\centering
\label{algori4}
  %\centering
  \begin{tabularx}{\textwidth}{lX}   
    Step 1 & With data $\{X_{0},\ldots,X_{T}\}$, construct the estimators $\widehat{m}_h(x)$ and $\widehat{\sigma}_h(x)$ by formula \cref{truncatedest}. Furthermore, compute predictive residuals based on \cref{preres}. Denote the empirical distribution of centered residuals $\hat{\epsilon}^p_t - \frac{1}{T}\sum_{i=1}^{T}\hat{\epsilon}^p_i, t= 1,\ldots,T$ by $\widehat{F}^p_{\epsilon}$.\\
    Steps 3-4 &  Same as in \cref{algori3} but
change the residual distribution from $\widehat{F}_{\epsilon}$ to $\widehat{F}^{p}_{\epsilon}$, and change the application of \cref{algori1} to \cref{algori2}.
  \end{tabularx}
\end{algorithm}

\FloatBarrier
\section{Asymptotic properties}\label{Sec:asymptotic}
In this section, we provide the theoretical substantiation for our non-parametric bootstrap prediction methods---\cref{algori1,algori2,algori3,algori4}. We start by analyzing optimal point predictions and QPI based on \cref{algori1,algori2}. 

\begin{Remarknn}
Since the effect of leaving out
one data pair $X_t \text{vs}~\{X_{t-1}\}$ is asymptotically negligible for large $T$, the delete-$X_t$ estimator $\widehat{m}^{t}(x)$ and $\widehat{\sigma}^{t}(x)$ are asymptotically equal to $\widehat{m}(x)$ and $\widehat{\sigma}(x)$, respectively. Then, the predictive residual $\hat{\epsilon}_t^{p}$ is asymptotically the same as the fitted residual $\hat{\epsilon}_{t}$; see Lemma 5.5 of \cite{pan2016bootstrap} for a formal comparison of these two types of estimators and residuals. Thus, we just give theorems to guarantee the asymptotic properties of point predictions and PIs with fitted residuals. The asymptotic properties for variants with predictive residuals also stand true. 
\end{Remarknn}
 
\subsection{On   point prediction and QPI}
First, to conduct statistical inference for time series, we need to quantify the degree of asymptotic dependence of time series. In this paper, we consider that the time series is geometrically ergodic which is equivalent to the $\beta$-mixing condition with  an exponentially fast mixing rate; see \cite{bradley2005basic} for a detailed introduction of different mixing conditions and ergodicity. To simplify the proof, we make the  following assumptions:
\begin{itemize}
    \item [A1] $|m(x)| + \sigma(x)\mathbb{E}|\epsilon_1| \leq c_1 + c_2|x|$ for all $x\in\mathbb{R}$ and some $c_1<\infty, c_2<1$;
    \item [A2] $\sigma(x)\geq c_3>0$ for all $x\in\mathbb{R}$ and some $c_3>0$; 
    \item [A3] $f_{\epsilon}(x)$ is positive everywhere.
\end{itemize}
A1---A3 can guarantee that the time series process is geometrically ergodic; see Theorem 1 of \cite{franke2004bootstrapping} for proof and see the work of \cite{min1999probabilistic} for a discussion on sufficient conditions of higher order time series. 

Since we need to build consistent properties of non-parametric estimation, we further assume:
\begin{itemize}
    \item [A4] For the regression function $m(x)$: it is twice continuously differentiable with bounded derivatives and we denote its Lipschitz continuous constant as $L_{m}$;
    \item [A5] For the volatility function $\sigma(x)$: it is twice continuously differentiable with bounded derivatives and we denote its Lipschitz continuous constant as $L_{\sigma}$. Moreover, for all $M<\infty$, there are $c_{M}<\infty$ with $\mathbb{E}|\sigma(X_0)\epsilon_1|^{M}\leq c_{M}$, where $X_0$ is the initial point of the time series;
    \item[A6] For $L_{m}$ and $L_{\sigma}$, we assume $L_{m} + L_{\sigma}\mathbb{E}|\epsilon_1|<1$;
    \item [A7] For the innovation distribution: % beyond A3,
 $f_{\epsilon}$ is twice continuously differentiable; $f_{\epsilon}, f^{'}_{\epsilon}$ and $f^{''}_{\epsilon}$ are bounded; and $\sup_{x\in\mathbb{R}}|xf^{'}_{\epsilon}(x)|<\infty$; 
    \item[A8] For the kernel function $K(x)$: it is a compactly supported and symmetric probability density on $\mathbb{R}$, and has bounded derivative.
\end{itemize}
\begin{Remarknn}
    The assumption A6 is originally used to show the expected value of $X_{t}^*$ is $O_{p}(1)$ in the bootstrap world for all $t$. In practice, this assumption is not strict; see examples in \cref{Sec:simulation}. For assumption A8, we can apply a kernel with a support on the whole real line as long as the part outside a large enough compact set is asymptotically negligible. 
\end{Remarknn}
Under A1---A8, \cite{franke2002bootstrap} show that truncated local constant estimators \cref{truncatedest} are uniformly consistent with the true functions in an expanding region. We summarize this result in the lemma below:
\begin{Lemma}\label{Lemma:uniformest}
Under A1---A8 and observed data $\{X_0,\ldots,X_T\}$, for local constant estimation as in  \cref{truncatedest}, we have:
\begin{equation}
         \sup_{|x|\leq c_{T}}|\widehat{m}_h(x) - m(x)|\overset{p}{\to}0~\text{and}~\sup_{|x|\leq c_{T}}|\widehat{\sigma}_h(x) - \sigma(x)|\overset{p}{\to}0.
\end{equation}
 
\noindent
where $c_T$ is an appropriate sequence that converges to infinity as $T\to \infty .$ 
\end{Lemma} 

  In addition, for the centered empirical distribution of $\hat{\epsilon}$, we can derive \cref{Lemma:uniformres} to describe its consistency property:
\begin{Lemma}\label{Lemma:uniformres}
Under A1---A8 and observed data $\{X_0,\ldots,X_T\}$, for the centered empirical distribution $\widehat{F}_{\epsilon}$, we have:
\begin{equation}\label{resuniform}
        \sup_{x\in\mathbb{R}}|\widehat{F}_{\epsilon}(x) - F_{\epsilon}(x)|\overset{p}{\to} 0.
\end{equation}
\end{Lemma}
See Theorem 5 of \cite{franke2002bootstrap} for the proof of \cref{Lemma:uniformest} and \cref{Lemma:uniformres}. Combining all pieces, we present \cref{Theorem:QPI} to show that optimal point prediction and QPI returned by \cref{algori1} or \cref{algori2} are consistent and asymptotically valid respectively, conditionally on the latest observations.

\begin{Theorem}\label{Theorem:QPI}
Under assumptions A1---A8 and observed data $\{X_0,\ldots,X_T\}$, we have:
\begin{equation}\label{claim1}
    \sup_{|x|\leq c_{T}}\left| F_{X^*_{T+k}|X_T,\ldots,X_0}(x) -  F_{X_{T+k}|X_T}(x)\right| \overset{p}{\to} 0,~\text{for}~k \geq 1,\end{equation}
where $X_{T+k}^*$ is the future value in the bootstrap world that can be determined iteratively by applying the expression $X_{T+i}^{*} =  \widehat{m}_{h}(X_{T+i-1}^*) + \widehat{\sigma}_{h}(X_{T+i-1}^*)\hat{\epsilon}^*_{T+i}$ for $i = 1,\ldots k$; $\{\hat{\epsilon}^*_{T+i}\}_{i=1}^{k}$ are i.i.d. with distribution given by the empirical distribution of fitted
(or predictive) residuals; $F_{X^*_{T+k}|X_T,\ldots,X_0}(x)$ represents the distribution
$\mathbb{P}^{*}(X^{*}_{T+k}\leq x|X_T,\ldots,X_0)$; here we take $\mathbb{P}^{*}$ to represent the probability measure conditional on sample of data; $F_{X_{T+k}|X_T}(x)$ represents the (conditional) distribution of $X_{T+k}$ in the real world, i.e., $\mathbb{P}(X_{T+k}\leq x|X_T)$.
\end{Theorem}

\subsection{On PPI with homoscedastic errors}\label{Subsec:PPI}
With more complicated prediction procedures such as \cref{algori3,algori4}, we expect to find a more accurate PI, i.e., a PPI. The superiority of such   PIs is that the estimation variability can be captured when we use the distribution of the predictive root in the bootstrap world to approximate its variant in the real world. We  consider models with homoscedastic errors throughout this section; the model with heteroscedastic errors will be analyzed later. 

Firstly, let us consider the one-step ahead predictive root centered at optimal $L_2$ point prediction in the real and bootstrap world as given below:
\begin{equation}\label{onesteppreroot}
\begin{split}
     &X_{T+1} - \widehat{X}_{T+1} = m(X_T)+\epsilon_{T+1} - \frac{1}{M}\sum_{i=1}^{M}\left(\widehat{m}_{h}(X_T) + \hat{\epsilon}_{i,T+1}\right); \\
     &X^*_{T+1} - \widehat{X}^*_{T+1} = \widehat{m}_{g}(X_T)+\hat{\epsilon}_{T+1}  - \frac{1}{M}\sum_{i=1}^{M}\left(\widehat{m}^*_{h}(X_T) + \hat{\epsilon}^*_{i,T+1}\right),
\end{split}
\end{equation}
where $M$ is the number of bootstrap replications we employ to approximate the optimal $L_2$ point prediction. Since we have centered the residuals to   mean zero, \cref{onesteppreroot} degenerates to the below simple form asymptotically as $M\to \infty$:
\begin{equation}\label{onestepprerootsimp}
\begin{split}
     &X_{T+1} - \widehat{X}_{T+1} = m(X_T)+\epsilon_{T+1} - \widehat{m}_{h}(X_T); \\
     &X^*_{T+1} - \widehat{X}^*_{T+1} = \widehat{m}_{g}(X_T)+\hat{\epsilon}_{T+1}  - \widehat{m}^*_{h}(X_T).
\end{split}
\end{equation}
To acquire a pertinent PI according to Definition 2.4 of \cite{pan2016bootstrap}, in addition to \cref{resuniform}, we also need asymptotically valid confidence intervals for local constant estimation in the bootstrap world, i.e., we 
should be able to estimate the distribution of the non-parametric estimator in the bootstrap world. For one-step ahead prediction, this condition can be formulated as below:
\begin{equation}\label{CIboot}
    \sup_{x}|\mathbb{P}(a_TA_{m}\leq x) - \mathbb{P}^{*}(a_TA^*_{m}\leq x)|\overset{p}{\to} 0,
\end{equation}
where 
\begin{equation}
\begin{split}
A_{m} = m(X_T) - \widehat{m}_h(X_T)~;~A^*_{m} = \widehat{m}_g(X_T) -  \widehat{m}^*_h(X_T)  \\
\end{split}
\end{equation}
\noindent 
and $a_T$ is an   appropriate sequence such that
$\mathbb{P}(a_TA_{m}\leq x)$ has a nontrivial limit 
as $T\to \infty$.   
%\begin{Remark}
\cite{pan2016bootstrap} assumed that the nontrivial limit of $\mathbb{P}(a_TA_{m}\leq x)$ is continuous. In this case,  the uniform convergence in \cref{CIboot} follows from the pointwise convergence of all $x$. 
%\end{Remark}

\begin{Remarknn}
As we have discussed in \cref{Remark:suboprate}, the bootstrap procedure can not capture the bias term of non-parametric estimation exactly unless with delicate manipulations. \cite{pan2016bootstrap} take two strategies to solve this issue: (B1) Let $g = h$ and take a bandwidth rate satisfying $hT^{1/5}\to 0$,  i.e., under-smoothing in function estimation; (B2) Use the optimal
smoothing rate with $h$
 proportional to $T^{-1/5}$, but generate time series in the bootstrap world with over-smoothing estimators, i.e., $g\neq h$ and $g/h\overset{}{\to}\infty$. No matter which approach we take, \cref{CIboot} can be shown; see details from Theorem 1 of \cite{franke2002bootstrap} and Theorem 5.4 of \cite{pan2016bootstrap}. 
\end{Remarknn}

The following corollary is immediate:
\begin{Corollary}\label{Corollary:1steppertinent}
Under assumptions A1---A3 and observed data $\{X_0,\ldots,X_T\}$, the one-step ahead PI returned by \cref{algori3} and \cref{algori4} with fitted or predictive residuals are asymptotically pertinent, respectively. 
\end{Corollary}

However, for multi-step ahead predictions, the analysis gets more complicated and the under-smoothing strategy turns out to work better. For example, considering the two-step ahead prediction, the two predictive roots can be written as below:
\begin{equation}\label{pretoot2stepreal}
\begin{split}
    X_{T+2} - \widehat{X}_{T+2} &= m(X_{T+1}) + \epsilon_{T+2} - \frac{1}{M}\sum_{i=1}^{M}\left(  \widehat{m}_{h}\left(\widehat{m}_{h}(X_T)+\hat{\epsilon}_{i,T+1}\right)+\hat{\epsilon}_{i,T+2}
 \right) \\
  &\approx m(m(X_T)+\epsilon_{T+1}) + \epsilon_{T+2} - \frac{1}{M}\sum_{i=1}^{M} \widehat{m}_{h}\left(\widehat{m}_{h}(X_T)+\hat{\epsilon}_{i,T+1}\right) .
\end{split}
\end{equation}
Correspondingly, the predictive root in the bootstrap world is:
\begin{equation}\label{pretoot2stepboot}
\begin{split}
    X^*_{T+2} - \widehat{X}^*_{T+2} &= \widehat{m}_{g}(X^*_{T+1}) + \hat{\epsilon}^*_{T+2} - \frac{1}{M}\sum_{i=1}^{M}\left(  \widehat{m}^*_{h}\left(\widehat{m}^*_{h}(X_T)+\hat{\epsilon}^*_{i,T+1}\right)+\hat{\epsilon}^*_{i,T+2}
 \right) \\
  &\approx \widehat{m}_{g}(\widehat{m}_{g}(X_T)+\hat{\epsilon}^*_{T+1}) + \hat{\epsilon}^*_{T+2} - \frac{1}{M}\sum_{i=1}^{M}  \widehat{m}^*_{h}\left(\widehat{m}^*_{h}(X_T)+\hat{\epsilon}^*_{i,T+1}\right),
\end{split}
\end{equation}
where the approximated equality is due to applying the LLN on the sample mean of centered residuals. 

\begin{Remarknn}\label{Remark:underopbandwidthdiss}
    We should notice that the over-smoothing approach may work better in finite samples. The reason is that applying the optimal bandwidth rate is superior when the bias-type term of the non-parametric estimation can be captured by the bootstrap. However, we will show soon that applying an under-smoothing bandwidth strategy is more accurate for multi-step ahead predictions since it can solve the bias issue and render a PPI. Thus, in practice, we recommend taking strategy (B2) to do one-step ahead predictions and taking strategy (B1) to do multi-step ahead predictions. For the time series with heteroscedastic errors, the optimal bandwidth strategy is slightly different; see \cref{Subsec:PPIHeter} for reference. 
\end{Remarknn}

Based on \cref{pretoot2stepreal,pretoot2stepboot}, as we prove the future distribution of $X_{T+k}^*$ converges uniformly to the future distribution of $X_{T+k}$ in probability, we can show the distribution of the predictive root $ X^*_{T+2} - \widehat{X}^*_{T+2}$ in the bootstrap world also converges uniformly in probability to the distribution of predictive root $X_{T+2} - \widehat{X}_{T+2}$ in the real world. This result guarantees the asymptotic validity of PPI. We summarize this conclusion in \cref{Theorem:PPI_part1}:

\begin{Theorem}\label{Theorem:PPI_part1}
    Under assumptions A1---A8 and observed data $\{X_0,\ldots,X_T\}$, we have:
\begin{equation}\label{claim2}
    \sup_{|x|\leq c_{T}}\left| F_{X^*_{T+k}- \widehat{X}^*_{T+k}|X_T,\ldots,X_0}(x) -  F_{X_{T+k} - \widehat{X}_{T+k}|X_T,\ldots,X_0}(x)\right| \overset{p}{\to} 0,~\text{for}~k \geq 1,\end{equation}
where $X_{T+k}^*-\widehat{X}^*_{T+k}$ is the k-step ahead predictive root in the bootstrap world and $F_{X^*_{T+k}- \widehat{X}^*_{T+k}|X_T,\ldots,X_0}(x)$ represents its distribution at point $x$; $X_{T+k} - \widehat{X}_{T+k}$ is the k-step ahead predictive root in the real world and $F_{X_{T+k} - \widehat{X}_{T+k}|X_T,\ldots,X_0}(x)$ represents its (conditional) distribution at point $x$. This theorem holds for both bandwidth selection strategies. 
\end{Theorem}

However, since we apply a more complicated procedure to capture 
estimation variability, 
we anticipate resulting in a PPI.  To see this, we first apply the Taylor expansion on the r.h.s. of \cref{pretoot2stepreal,pretoot2stepboot}; the two predictive roots can be decomposed into several parts:
\begin{equation}\label{2prerootaftertaylor}
\begin{split}
    &X_{T+2} - \widehat{X}_{T+2}  = m(m(X_T)) - \widehat{m}_{h}(\widehat{m}_{h}(X_T)) + m^{(1)}(\hat{x})\epsilon_{T+1} + \epsilon_{T+2}  - \frac{1}{M}\sum_{i=1}^{M}\widehat{m}^{(1)}_{h}(\Hat{\Hat{x}}_i)\hat{\epsilon}_{i,T+1};\\
    & X^*_{T+2} - \widehat{X}^*_{T+2} = \widehat{m}_{g}(\widehat{m}_{g}(X_T)) - \widehat{m}^*_{h}(\widehat{m}^*_{h}(X_T)) + \widehat{m}_{g}^{(1)}(\hat{x}^*)\hat{\epsilon}^*_{T+1} + \hat{\epsilon}^*_{T+2}  - \frac{1}{M}\sum_{i=1}^{M}\widehat{m}^{*(1)}_{h}(\Hat{\Hat{x}}_i^*)\hat{\epsilon}^*_{i,T+1},
\end{split}
\end{equation}
where $\hat{x}$ and $\hat{x}^*$ are some points between $m(X_T)$ and $m(X_T)+\epsilon_{T+1}$, $\widehat{m}_{g}(X_T)$ and $\widehat{m}_{g}(X_T) + \hat{\epsilon}_{T+1}^*$, respectively; $\Hat{\Hat{x}}_i$ and $\Hat{\Hat{x}}_i^*$ are some points between $\widehat{m}_{h}(X_T)$ and $\widehat{m}_{h}(X_T)+\hat{\epsilon}_{i,T+1}$, $\widehat{m}^*_{h}(X_T)$ and $\widehat{m}^*_{h}(X_T)+\hat{\epsilon}^*_{i,T+1}$, respectively; $k$-step ahead predictive root can be expressed similarly when $k>2$. We can think the r.h.s of \cref{2prerootaftertaylor} is made up of two components in both real and bootstrap worlds: (1) The two-step ahead estimation variability component, $m(m(X_T)) - \widehat{m}_{h}(\widehat{m}_{h}(X_T))$ and $\widehat{m}_{g}(\widehat{m}_{g}(X_T)) -  \widehat{m}^*_{h}(\widehat{m}^*_{h}(X_T))$; (2) The rest of terms, which are related to future innovations. For the second component, the bootstrap can mimic the real-world situation well.

We expect that the first component, i.e., variability of local constant estimation on the mean function $m(m(X_T)) - \widehat{m}_{h}(\widehat{m}_{h}(X_T))$ can be well approximated by its variant $\widehat{m}_{g}(\widehat{m}_{g}(X_T)) -  \widehat{m}^*_{h}(\widehat{m}^*_{h}(X_T))$ in the bootstrap world. Although PPIs with either of the  two bandwidth selection approaches are both asymptotically valid, the PPI with bandwidth strategy (B2) is only ``almost'' pertinent for multi-step ahead predictions since the variability of local constant estimation is not well estimated in finite samples; see also the simulation results in \cref{Sec:simulation} and  \hyperref[Appendix:diffoverunderband]{Appendix C}.  On the other hand, the PPI with the bandwidth strategy (B1) meets our goal. We summarize this finding in \cref{Theorem:PPI_part2}. 
\begin{Theorem}\label{Theorem:PPI_part2}
    Under assumptions A1---A8 and with observed data $\{X_0,\ldots,X_T\}\in \Omega_T$, where $\mathbb{P}((X_0,\ldots,X_T)\in \Omega_T) = 1 - o(1)$ as $T\to \infty$, by taking the bandwidth strategy (B1), we can build confidence bound for the local constant estimation at $k$-step:
\begin{equation}\label{kstepestdist}
\begin{split}
    \sup_{|x|\leq c_{T}} & \left| \mathbb{P}\left( a_T\left(\mathcal{M}_{k}(X_T) - \widehat{\mathcal{M}}_{h,k}(X_T)   \right)\leq x  \right) -  \right. \\
    &\left. \mathbb{P}\left( a_T\left(\mathcal{M}^*_{h,k}(X_T) - \widehat{\mathcal{M}}_{h,k}^*(X_T)   \right)\leq x  \right)
     \right| \overset{p}{\to} 0,~\text{for}~k \geq 1;
\end{split}
\end{equation}
$\mathcal{M}_{k}(X_T)$ can be expressed by computing $X_{T+i} = m(X_{T+i-1})$ iteratively for $i = 1,\ldots, k$, i.e., it has the  form   below:
\begin{equation}
\mathcal{M}_{k}(X_T) = m(m(\cdots (m(m(X_T)))));
\end{equation}
$\widehat{\mathcal{M}}_{h,k}(X_T) $ can be expressed by computing $X_{T+i} = \widehat{m}_{h}(X_{T+i-1})$ iteratively for $i = 1,\ldots, k$, i.e., it has the form  below:
\begin{equation}
\widehat{\mathcal{M}}_{h,k}(X_T) = \widehat{m}_h(\widehat{m}_h(\cdots \widehat{m}_h(\widehat{m}_h(X_T))\cdots));
\end{equation}
$\mathcal{M}^*_{h,k}(X_T) $ and $\widehat{\mathcal{M}}_{h,k}^*(X_T)$ can be expressed similarly. 
\end{Theorem}
The direct implication of \cref{Theorem:PPI_part2} is that the PPI generated by \cref{algori3} and \cref{algori4} should have better CVR for small sample sizes than QPI since the estimation variability is included in the PI with \textit{high probability}; see the simulation examples in \cref{Sec:simulation}.

\subsection{On PPI with heteroscedastic errors}\label{Subsec:PPIHeter}
For time series models with heteroscedastic errors, i.e., where the variance function $\sigma(x)$ represents the heteroscedasticity of innovations, we may not need to care about the bias  term in the non-parametric estimation of variance function. In other words, we neither use under-smoothing nor over-smoothing bandwidth tricks on the variance function to generate the bootstrap series for covering the bias term; we can just use the bandwidth with optimal rate to estimate the variance function from real and bootstrap series. 

To see this, let us  consider the two-step ahead predictive root with heteroscedastic errors. In the real world, we have:
\begin{equation}\label{pretoot2steprealheter}
\begin{split}
    X_{T+2} - \widehat{X}_{T+2} &= m(X_{T+1}) + \sigma(X_{T+1})\epsilon_{T+2} - \frac{1}{M}\sum_{i=1}^{M}\left(  \widehat{m}_{h}\left(\widehat{m}_{h}(X_T)+\widehat{\sigma}_h(X_T)\hat{\epsilon}_{i,T+1}\right)+\widehat{\sigma}_h(X_{T+1})\hat{\epsilon}_{i,T+2}
 \right) \\
  &\approx m(m(X_T)+\sigma(X_{T})\epsilon_{T+1}) + \sigma(X_{T+1})\epsilon_{T+2} - \frac{1}{M}\sum_{i=1}^{M} \widehat{m}_{h}\left(\widehat{m}_{h}(X_T)+\widehat{\sigma}_h(X_T)\hat{\epsilon}_{i,T+1}\right).
\end{split}
\end{equation}
Correspondingly, the predictive root in the bootstrap world is:
\begin{equation}\label{pretoot2stepbootheter}
\begin{split}
    X^*_{T+2} - \widehat{X}^*_{T+2} &= \widehat{m}_{g}(X^*_{T+1}) + \widehat{\sigma}_{g}(X_{T+1}^*){\epsilon}^*_{T+2} - \frac{1}{M}\sum_{i=1}^{M}\left(  \widehat{m}^*_{h}\left(\widehat{m}^*_{h}(X_T)+\widehat{\sigma}_{h}^{*}(X_T)\hat{\epsilon}^*_{i,T+1}\right)+\widehat{\sigma}_{h}^{*}(X^*_{T+1})\hat{\epsilon}^*_{i,T+2}
 \right) \\
  &\approx \widehat{m}_{g}(\widehat{m}_{g}(X_T)+\widehat{\sigma}_{g}(X_{T}^*)\hat{\epsilon}^*_{T+1}) + \widehat{\sigma}_{g}(X_{T+1}^*)\hat{\epsilon}^*_{T+2} - \frac{1}{M}\sum_{i=1}^{M}  \widehat{m}^*_{h}\left(\widehat{m}^*_{h}(X_T)+\widehat{\sigma}_{h}^{*}(X_T)\hat{\epsilon}^*_{i,T+1}\right).
\end{split}
\end{equation}
Through Taylor expansion, we can get:
\begin{equation}\label{2prerootaftertaylorheter}
\begin{split}
    X_{T+2} - \widehat{X}_{T+2}  &\approx  m(m(X_T)) - \widehat{m}_{h}(\widehat{m}_{h}(X_T)) \\
    &+ m^{(1)}(\hat{x})\sigma(X_{T})\epsilon_{T+1} + \sigma(X_{T+1})\epsilon_{T+2}  - \frac{1}{M}\sum_{i=1}^{M}\widehat{m}^{(1)}_{h}(\Hat{\Hat{x}}_i)\widehat{\sigma}_h(X_T)\hat{\epsilon}_{i,T+1};\\
    X^*_{T+2} - \widehat{X}^*_{T+2} &\approx  \widehat{m}_{g}(\widehat{m}_{g}(X_T)) - \widehat{m}^*_{h}(\widehat{m}^*_{h}(X_T)) \\
    &+ \widehat{m}_{g}^{(1)}(\hat{x}^*)\widehat{\sigma}_{g}(X_{T}^*)\hat{\epsilon}^*_{T+1} + \widehat{\sigma}_{g}(X_{T+1}^*)\hat{\epsilon}^*_{T+2}  - \frac{1}{M}\sum_{i=1}^{M}\widehat{m}^{*(1)}_{h}(\Hat{\Hat{x}}_i^*)\widehat{\sigma}_{h}^{*}(X_T)\hat{\epsilon}^*_{i,T+1}.
\end{split}
\end{equation}
We can still think that the r.h.s. of \cref{2prerootaftertaylorheter} contains two components. Once we use the under-smoothing technique to cover the estimation variability for the mean function, since the residual distribution is determined by the estimated mean and variance functions, the convergence rate of residual distribution to the true innovation distribution is dominated by the convergence rate of $\widehat{m}_h(x)$ to $m(x)$. In addition, all estimators of the variance function in \cref{2prerootaftertaylorheter} are tied with future estimated innovations; so we are free to use the bandwidth $g=h$ with optimal
smoothing rate to estimate the variance function, and the overall convergence rate will not change. To show this benefit, we run some simulations in \hyperref[Appendix:optbandwidthonestvar]{Appendix D} to compare the performance of PIs   applying under-smoothing or optimal bandwidth on estimating the variance function. In \cref{Sec:simulation}, we will take optimal bandwidth to estimate the variance function if the time series is heteroscedastic.  

To analyze the pertinence of PPI for time series with heteroscedastic errors, from \cref{2prerootaftertaylorheter} it is apparent that the distribution of $m(m(X_T)) - \widehat{m}_{h}(\widehat{m}_{h}(X_T))$ can still be approximated by $\widehat{m}_{g}(\widehat{m}_{g}(X_T)) - \widehat{m}^*_{h}(\widehat{m}^*_{h}(X_T))$. For the rest of the terms, the bootstrap can still mimic the real-world situation.

\section{Simulations}\label{Sec:simulation}
In this section, we deploy simulations to check the performance of 5-step ahead point predictions and corresponding PIs of our algorithms in the $R$ platform with finite samples. To get the optimal bandwidth $h_{op}$ for our local constant estimators, we rely on the function \textit{npregbw} from the $R$ package \textit{np}. For the under-smoothing and over-smoothing bandwidth, we take them as $0.5\cdot h_{op}$ and $2\cdot h_{op}$, respectively.

\subsection{Optimal point prediction}
We first consider a simple non-linear model:
\begin{equation}\label{model1}
    X_t = \log(X_{t-1}^2 + 1) + \epsilon_t,
\end{equation}
where $\{\epsilon_t\}$ are assumed to have a standard normal distribution. The geometric ergodicity of \cref{model1} can be easily checked. 

We apply the ``oracle'' prediction to be the benchmark. The oracle prediction is returned by a simulation approach assuming we know the true model and the error distribution, i.e., the simulation-based prediction as we discussed in \cref{Sec:Intro}; see Section 3.2 of \cite{wu2023bootstrap} for more details and theoretical validation about this approach. Since this oracle prediction should stand for the best performance, we would like to challenge our non-parametric bootstrap-based methods by comparing them with the oracle prediction. We also pretend that the true model and innovation distribution are unknown when we do the non-parametric bootstrap-based prediction. For point predictions, we just utilize fitted residuals. The application of predictive residuals will play a role in building PIs later.

In a single experiment, we take $X_0\sim \text{Uniform}(-1,1)$, then generate a series with size $C+T+1$ according to \cref{model1} iteratively. Here, $C$ is taken as $200$ to remove the effects of the initial distribution of $X_0$. To perform oracle predictions, we take $M = 1000$ to get a satisfying approximation. For a fair comparison, we also do
$1000$ times bootstrap in \cref{algori1,algori2} to get bootstrap-based predictions. 

Referring to the simulation studies of \cite{pan2016bootstrap}, we take $T=100,~200$, $k=1,\ldots,5$ and deploy the Mean Squared Prediction Error (MSPE) to compare oracle and bootstrap predictions. The metric MSPE can be approximated based on the below formula:
\begin{equation}
    \text{MSPE of the}~k \text{-th ahead prediction} = \frac{1}{N}\sum_{n=1}^{N}(X_{n,k} - P_{n,k})^2,~\text{for}~k=1,\ldots,5,
\end{equation}
where $P_{n,k}$ represents the $k$-th step ahead optimal $L_1$ or $L_2$ point predictions implied by the bootstrap or simulation approach, and $X_{n,k}$ stands for the true future value in the $n$-th replication. We take $N = 5000$ and record all MSPEs in \cref{Tablogx2add1_point}.
\begin{table}[htbp]
\centering
  \caption{The MSPEs of different predictions under Model \cref{model1} with a standard normal innovation}
  \vspace{2pt}
  \label{Tablogx2add1_point}
\begin{tabular}{llccccc}
  \toprule 
 Model: & \multicolumn{6}{c}{$X_t = \log(X_{t-1}^2 + 1) + \epsilon_t, \epsilon_t\sim N(0,1)$} \\
 \midrule
    $T = 100$ & Prediction step & 1     & 2     & 3     & 4     & 5 \\[3pt]
    $L_2$-Bootstrap &     & 1.1088 & 1.5223 & 1.6088 & 1.5886 & 1.6282 \\
    $L_1$-Bootstrap &    & 1.1123 & 1.5290 & 1.6212 & 1.6011 & 1.6385 \\
    $L_2$-Oracle &  & 1.0181 & 1.4521 & 1.5529 & 1.5273 & 1.5731 \\
    $L_1$-Oracle &   & 1.0198 & 1.4540 & 1.5554 & 1.5305 & 1.5734 \\[3pt]
    $T = 200$ &       &       &       &       &       &  \\[3pt]
    $L_2$-Bootstrap &     & 1.0142 & 1.4006 & 1.5380 & 1.5956 & 1.6102\\
    $L_1$-Bootstrap &    & 1.0134 & 1.4041 & 1.5426 & 1.6024 & 1.6171  \\
    $L_2$-Oracle &  & 0.9790 & 1.3671 & 1.4982 & 1.5556 & 1.5791 \\
    $L_1$-Oracle &   & 0.9793 & 1.3681 & 1.4999 & 1.5568 & 1.5791 \\
       \bottomrule
    \end{tabular}\\
    %   \tiny
    %   \raggedright
    %  \textit{Note:} 
\end{table}

From \cref{Tablogx2add1_point}, we can find that MSPEs of oracle and bootstrap-based $L_1$ or $L_2$ optimal predictions are very close to each other, respectively. MSPEs of oracle optimal predictions are always smaller than corresponding bootstrap predictions. This phenomenon is in our expectation since the bootstrap prediction is obtained with an estimated model and innovation distribution. 

Rather than applying the standard normal distribution, we consider a skewed innovation, i.e., $\epsilon_t\sim\chi^2(3)-3$. Repeat the above process, we present MSPEs in \cref{Tablogx2add1_point_chi}.
\begin{table}[htbp]
\centering
  \caption{The MSPEs of different predictions under Model \cref{model1} with $\chi(3)-3$ innovation}
  \vspace{2pt}
  \label{Tablogx2add1_point_chi}
\begin{tabular}{llccccc}
  \toprule 
 Model: & \multicolumn{6}{c}{$X_t = \log(X_{t-1}^2 + 1) + \epsilon_t, \epsilon_t\sim\chi(3)-3 $} \\
 \midrule
    $T = 100$ & Prediction step & 1     & 2     & 3     & 4     & 5 \\[3pt]
    $L_2$-Bootstrap &     & 6.7286 & 7.6087 & 7.8202 & 7.3395 & 7.6966  \\
    $L_1$-Bootstrap &    & 7.1093 & 7.9908 & 8.2598 & 7.6761 & 7.9988\\
    $L_2$-Oracle &  & 6.2972 & 7.3608 & 7.6953 & 7.1766 & 7.5157 \\
    $L_1$-Oracle &   & 6.6937 & 7.6540 & 8.0064 & 7.3889 & 7.7174   \\[3pt]
    $T = 200$ &       &       &       &       &       &  \\[3pt]
    $L_2$-Bootstrap &     & 6.2457 & 7.1662 & 7.5042 & 7.6227 & 7.1980\\
    $L_1$-Bootstrap &    & 6.6355 & 7.4942 & 7.7964 & 7.9285 & 7.5006  \\
    $L_2$-Oracle &  &  5.9531 & 7.0244 & 7.3823 & 7.4382 & 7.0738  \\
    $L_1$-Oracle &   & 6.3519 & 7.2785 & 7.5810 & 7.6443 & 7.2600 \\
       \bottomrule
    \end{tabular}\\
    %   \tiny
    %   \raggedright
    %  \textit{Note:} 
\end{table}
The performance of bootstrap-based predictions is also competitive with oracle predictions. Another notable phenomenon indicated by \cref{Tablogx2add1_point_chi} is that the MSPE of $L_2$ optimal prediction is always less than its corresponding $L_1$ optimal prediction. The reason for this is that the $L_2$ optimal prediction coincides with the $L_2$ loss used in MSPE. However, this phenomenon is not remarkable for results in \cref{Tablogx2add1_point} since the innovation distribution is symmetric in that case.

For the non-linear model with heteroscedastic errors, we consider the following model:
\begin{equation}\label{model2}
    X_t = \sin(X_{t-1}) + \epsilon_t\sqrt{0.5 + 0.25X_{t-1}^2}.
\end{equation}
The Model \cref{model2} is in a GARCH form except that the regression function is non-linear. This model was also considered by \cite{pan2016bootstrap}. We present MSPEs of different predictions in \cref{Tablogsinxaddsqrtx_point}. It reveals that our bootstrap-based optimal point prediction methods can work for the non-linear time series model with heteroscedastic errors and its performance is still competitive with oracle predictions. 

\begin{table}[H]
\centering
  \caption{The MSPEs of different predictions under Model \cref{model2} with standard normal innovation}
  \vspace{2pt}
  \label{Tablogsinxaddsqrtx_point}
\begin{tabular}{llccccc}
  \toprule 
 Model: & \multicolumn{6}{c}{$X_t = \sin(X_{t-1}) + \epsilon_t\sqrt{0.5 + 0.25X_{t-1}^2}, \epsilon_t\sim N(0,1) $} \\
 \midrule
    $T = 100$ & Prediction step & 1     & 2     & 3     & 4     & 5 \\[3pt]
    $L_2$-Bootstrap &     & 0.9447 & 1.1306 & 1.2373 & 1.2091 & 1.2714\\
    $L_1$-Bootstrap &    &  0.9461 & 1.1374 & 1.2396 & 1.2127 & 1.2731\\
    $L_2$-Oracle &  & 0.8454 & 1.0726 & 1.1832 & 1.1722 & 1.2186 \\
    $L_1$-Oracle &   & 0.8457 & 1.0730 & 1.1841 & 1.1737 & 1.2183  \\[3pt]
    $T = 200$ &       &       &       &       &       &  \\[3pt]
    $L_2$-Bootstrap &     & 0.8798 & 1.1539 & 1.2600 & 1.2901 & 1.2717  \\
    $L_1$-Bootstrap &    & 0.8833 & 1.1600 & 1.2649 & 1.2949 & 1.2749     \\
    $L_2$-Oracle &  & 0.8103 & 1.0991 & 1.2227 & 1.2680 & 1.2509 \\
    $L_1$-Oracle &   &  0.8107 & 1.1000 & 1.2239 & 1.2684 & 1.2511 \\
       \bottomrule
    \end{tabular}\\
    %   \tiny
    %   \raggedright
    %  \textit{Note:} 
\end{table}

\begin{Remark}
In practice, we should mention that both local constant estimators $\widehat{m}(x)$ and $\widehat{\sigma}(x)$ will only be accurate when $x$ falls in the area where data are dense. Estimations in the sparse area will return large fitted residuals. This large residual will spoil the multi-step ahead prediction process in the bootstrap procedure. %To see the reason, image that $\{\hat{\epsilon}^*_{T+1},\ldots,\hat{\epsilon}^*_{T+k}\}$ are all generated with large values, $\{X^{*}_{T+1},\ldots,X^*_{T+k} \}$ tend to become larger and larger. Then, the denominator of $\widehat{m}(x)$ and $\widehat{\sigma}(x)$ will be 0 numerically, so that we may get a NaN pseudo value $X^*_{T+k}$. Recall that we take the mean or median of pseudo values $\{X^{(i)}_{T+k} \}_{i=1}^{M}$ to be the optimal prediction of $X_{T+k}$, this NaN pseudo value will destroy this step. 
Thus, depending on which optimal prediction we are pursuing, we replace all inappropriate or numerical NaN values with the sample mean or sample median of observed data. In addition, during the simulation studies, we truncate $\widetilde{m}(x)_{h}$, i.e., we take $C_m$ as $5\cdot \max\{|x_0|,\ldots,|x_T|\}$. For the mean function estimator $\widetilde{m}(x)_{h}^*$ in the bootstrap world, we take $C^*_m$ as $\min\{2\cdot C_m, 5\cdot\max\{|x^*_0|,\ldots,|x^*_T|\}\}$ since we want to allow more variability for bootstrap series. For the local constant estimator of the variance function, we take $c_{\sigma}$ and $c^*_{\sigma}$ as $0.01$. We take $C_{\sigma}$ and $C^*_{\sigma}$ as $2\cdot \hat{\sigma}$ and $\min\{4\cdot \hat{\sigma},2\cdot \hat{\sigma}^*\}$, respectively; $\hat{\sigma}$ and $\hat{\sigma}^*$ are the sample standard deviations of observed series in the real world and bootstrap world, respectively. These truncating constants work well for the above two models. In practice, a cross-validation approach could be taken to find optimal truncating constants. 
\end{Remark}
\subsection{QPI and PPI}
In this subsection, we try to evaluate the CVR of QPI and PPI based on the non-parametric forward bootstrap prediction method. Similarly, we take the oracle prediction interval as the benchmark, which is computed by QPI with known model and innovation distribution; see discussion in \cref{Sec:Intro} and Section 3.2 of \cite{wu2023bootstrap} for references on this approach. 

% We thought the construction of PI is a more complicated problem and we try to perform simulation studies as comprehensively as we can, so we consider one more following Threshold model:
% \begin{equation}\label{model3}
%      X_t = (0.1\cdot X_{t-1}\cdot + 0.5\cdot e^{-X_{t-1}^2}\cdot\epsilon_t) I(X_{t-1}\leq 0) + (0.8\cdot X_{t-1} +\epsilon_t) I(X_{t-1}>0).
% \end{equation}
% In total, we consider three non-linear models---\cref{model1,model2,model3}. 

Due to the time complexity of double bootstrap in the bootstrap world, we only take $B = 500$ and $M = 100$ in \cref{algori3,algori4} to derive PPI. Correspondingly, we take $M = 500$ to compute the QPI. In practice, people can increase the value of $B$ and $M$. For getting a result as consistent as possible, we still repeat the simulation process 5000 times. 

The empirical CVR of bootstrap-based QPI and PPI for $k = 1,\ldots,5$ step ahead predictions with the below formula:
\begin{equation}\label{CVRformula}
    \text{CVR of the}~k \text{-th ahead prediction} = \frac{1}{N}\sum_{n=1}^{N}\mathbbm{1}_{X_{n,k}\in [L_{n,k},U_{n,k}]}, \text{for}~k=1,\ldots,5,
\end{equation}
where $[L_{n,k},U_{n,k}]$ and $X_{n,k}$ represent the $k$-th step ahead prediction interval and the true future value in the $n$-th replication, respectively. In addition to CVR, we are also concerned about the empirical LEN of different PIs. The empirical LEN of a PI is defined as below:
\begin{equation}
        \text{LEN of the}~k \text{-th ahead PI} = \frac{1}{N}\sum_{n=1}^{N}(U_{n,k}-L_{n,k}), \text{for}~k=1,\ldots,5.
\end{equation}
Recall that the PPI can be centered at $L_1$ or $L_2$ optimal point predictor and QPI can be found with optimal bandwidth and under-smoothing bandwidth, thus we have four types of PIs based on bootstrap. In particular, each type of PI can be performed with fitted or predictive residuals. We totally have 8 bootstrap-type PIs and one oracle PI. Besides, to observe the effects of introducing the predictive residuals and the superiority of PPI, we consider three sample sizes, 50, 100 and 200. All CVRs and LENs for different PIs on predicting \cref{model1} and \cref{model2} are presented in \cref{Tab:model1} and \cref{Tab:model2}, respectively. 

From there, we can observe that the SPI (oracle PI) is the best one according to the most accurate CVR and relatively small LEN. For the QPI with fitted residuals, it under-covers the true future value severely especially for data with a small sample size. With predictive residuals, although the LEN of PI gets amplified, the CVR of QPI improves significantly. After applying the under-smoothing bandwidth with QPI, the CVR further gets improved for multi-step ahead (i.e., $k\geq 2$) predictions no matter with fitted or predictive residuals. For PPI with fitted residuals, it outperforms the QPI with fitted residuals. For PPI with predictive residuals, it can achieve the most accurate CVR among various bootstrap-based PIs especially when the data is short, though the price is that its LEN is the largest compared to other PIs. We should notice that the QPI with predictive residuals and under-smoothing bandwidth can achieve great CVR with 200 samples for these two models. However, we may not know the sufficiently large sample size to guarantee that the QPI can work well. Thus, we will recommend taking the PPI with predictive residuals to be the first choice. 

\begin{Remark}
    We should clarify that the CVR computed by \cref{CVRformula} is the unconditional coverage rate of $X_{T+k}$ since it is an average of the conditional coverage of $X_{T+k}$ for all replications. 
\end{Remark}

\begin{table}[H]
\centering
  \caption{The CVR and LEN of PIs for \cref{model1}}
  \vspace{2pt}
  \label{Tab:model1}
\begin{tabular}{lcccccccccc}
  \toprule 
 Model 1: & \multicolumn{10}{c}{$X_t = \log(X_{t-1}^2 + 1) + \epsilon_t, \epsilon_t\sim N(0,1)$} \\
 \midrule
  & \multicolumn{5}{c}{CVR for each step} & \multicolumn{5}{c}{LEN for each step}\\
    $T = 200$  & 1     & 2     & 3     & 4     & 5 & 1 & 2 & 3 & 4 & 5  \\[3pt]
  QPI-f & 0.936 & 0.935 & 0.931 & 0.928 & 0.925 & 3.80 & 4.38 & 4.52 & 4.55 & 4.57 \\ 
  QPI-p & 0.943 & 0.944 & 0.939 & 0.935 & 0.937 & 3.94 & 4.54 & 4.69 & 4.73 & 4.74  \\ 
  QPI-f-u & 0.936 & 0.941 & 0.940 & 0.937 & 0.937 & 3.80 & 4.51 & 4.69 & 4.76 & 4.77 \\ 
  QPI-p-u & 0.942 & 0.949 & 0.949 & 0.945 & 0.949 & 3.95 & 4.68 & 4.86 & 4.92 & 4.94 \\ 
  $L_2$-PPI-f-u & 0.940 & 0.944 & 0.944 & 0.940 & 0.939 & 3.94 & 4.59 & 4.76 & 4.81 & 4.83 \\ 
  $L_2$-PPI-p-u & 0.947 & 0.954 & 0.951 & 0.947 & 0.947 & 4.09 & 4.75 & 4.92 & 4.98 & 5.00  \\ 
  $L_1$-PPI-f-u & 0.942 & 0.945 & 0.944 & 0.940 & 0.941 & 3.95 & 4.61 & 4.77 & 4.83 & 4.84 \\ 
  $L_1$-PPI-p-u & 0.948 & 0.954 & 0.952 & 0.948 & 0.949 & 4.10 & 4.77 & 4.94 & 4.99 & 5.01  \\ 
  SPI & 0.951 & 0.948 & 0.950 & 0.944 & 0.946 & 3.88 & 4.58 & 4.77 & 4.82 & 4.84 \\[3pt]
    $T = 100$        &       &       &       &       &  \\[3pt]
QPI-f & 0.921 & 0.918 & 0.912 & 0.913 & 0.909 & 3.74 & 4.28 & 4.40 & 4.44 & 4.45 \\
  QPI-p & 0.940 & 0.935 & 0.931 & 0.931 & 0.928 & 3.99 & 4.54 & 4.67 & 4.71 & 4.72 \\ 
  QPI-f-u & 0.916 & 0.928 & 0.931 & 0.930 & 0.927 & 3.74 & 4.46 & 4.63 & 4.69 & 4.71 \\ 
  QPI-p-u & 0.937 & 0.943 & 0.943 & 0.944 & 0.943 & 3.99 & 4.72 & 4.89 & 4.95 & 4.97 \\ 
  $L_2$-PPI-f-u & 0.931 & 0.934 & 0.935 & 0.934 & 0.931 & 3.97 & 4.58 & 4.73 & 4.78 & 4.80 \\ 
  $L_2$-PPI-p-u & 0.949 & 0.948 & 0.947 & 0.944 & 0.947 & 4.22 & 4.84 & 4.99 & 5.04 & 5.07 \\ 
  $L_1$-PPI-f-u & 0.931 & 0.936 & 0.934 & 0.933 & 0.934 & 3.98 & 4.60 & 4.75 & 4.79 & 4.82 \\ 
  $L_1$-PPI-p-u & 0.949 & 0.948 & 0.949 & 0.944 & 0.948 & 4.23 & 4.86 & 5.01 & 5.06 & 5.09 \\ 
  SPI & 0.951 & 0.941 & 0.946 & 0.942 & 0.944& 3.89 & 4.58 & 4.76 & 4.82 & 4.84  \\ [3pt]
    $T = 50$        &       &       &       &       &  \\[3pt]
QPI-f & 0.891 & 0.898 & 0.899 & 0.890 & 0.887 & 3.64 & 4.14 & 4.25 & 4.29 & 4.30  \\ 
  QPI-p & 0.923 & 0.926 & 0.931 & 0.924 & 0.917 & 4.04 & 4.56 & 4.67 & 4.71 & 4.72 \\ 
  QPI-f-u & 0.884 & 0.916 & 0.921 & 0.918 & 0.907 & 3.64 & 4.37 & 4.54 & 4.60 & 4.62 \\ 
  QPI-p-u & 0.914 & 0.939 & 0.940 & 0.939 & 0.934 & 4.03 & 4.79 & 4.95 & 5.00 & 5.02 \\ 
  $L_2$-PPI-f-u & 0.906 & 0.924 & 0.924 & 0.927 & 0.919 & 3.99 & 4.56 & 4.69 & 4.74 & 4.76 \\ 
  $L_2$-PPI-p-u & 0.936 & 0.951 & 0.948 & 0.944 & 0.943 & 4.41 & 4.97 & 5.10 & 5.15 & 5.16 \\ 
  $L_1$-PPI-f-u & 0.907 & 0.925 & 0.924 & 0.927 & 0.920 & 4.00 & 4.58 & 4.72 & 4.76 & 4.79 \\ 
  $L_1$-PPI-p-u & 0.939 & 0.952 & 0.948 & 0.945 & 0.941 & 4.43 & 5.00 & 5.12 & 5.17 & 5.18 \\ 
  SPI & 0.947 & 0.949 & 0.944 & 0.947 & 0.942 & 3.88 & 4.58 & 4.76 & 4.81 & 4.84 \\ 
       \bottomrule
    \end{tabular}\\
      \raggedright
     \textit{Note:} With no other specifications, throughout all simulations, QPI-f and QPI-p represent QPIs based on optimal bandwidth with fitted and predictive residuals, respectively; QPI-f-u and QPI-p-u represent QPIs based on under-smoothing bandwidth with fitted and predictive residuals, respectively; $L_2$-PPI-f-u and $L_2$-PPI-p-u represent PPIs centered at $L_2$ optimal point prediction with fitted and predictive residuals, respectively;  $L_1$-PPI-f-u and $L_1$-PPI-p-u represent PPIs centered at $L_1$ optimal point prediction with fitted and predictive residuals, respectively; All PPIs with ``-u'' symbol are based on applying under-smoothing bandwidth to estimate model; SPI represents the oracle PI. 
\end{table}

\begin{table}[htbp]
\centering
  \caption{The CVR and LEN of PIs for \cref{model2}}
  \vspace{2pt}
  \label{Tab:model2}
\begin{tabular}{lcccccccccc}
  \toprule 
 Model 2: & \multicolumn{10}{c}{$X_t = \sin(X_{t-1}) + \epsilon_t\sqrt{0.5 + 0.25X_{t-1}^2}, \epsilon_t\sim N(0,1)$} \\
 \midrule
  & \multicolumn{5}{c}{CVR for each step} & \multicolumn{5}{c}{LEN for each step}\\
    $T = 200$  & 1     & 2     & 3     & 4     & 5 & 1 & 2 & 3 & 4 & 5  \\[3pt]
 QPI-f & 0.913 & 0.918 & 0.916 & 0.924 & 0.924 & 3.30 & 3.93 & 4.07 & 4.11 & 4.12 \\ 
  QPI-p & 0.935 & 0.936 & 0.933 & 0.941 & 0.940 & 3.62 & 4.29 & 4.46 & 4.49 & 4.51  \\ 
  QPI-f-u & 0.904 & 0.934 & 0.935 & 0.943 & 0.944 & 3.34 & 4.25 & 4.50 & 4.55 & 4.57 \\ 
  QPI-p-u & 0.926 & 0.949 & 0.951 & 0.958 & 0.955 & 3.65 & 4.62 & 4.89 & 4.95 & 4.97  \\ 
  $L_2$-PPI-f-opv & 0.909 & 0.938 & 0.937 & 0.948 & 0.946 & 3.51 & 4.38 & 4.60 & 4.65 & 4.67 \\ 
  $L_2$-PPI-p-opv & 0.932 & 0.952 & 0.951 & 0.961 & 0.959 & 3.87 & 4.80 & 5.03 & 5.08 & 5.10 \\ 
  $L_1$-PPI-f-opv & 0.912 & 0.939 & 0.937 & 0.949 & 0.946 & 3.53 & 4.38 & 4.59 & 4.64 & 4.66 \\ 
  $L_1$-PPI-p-opv & 0.933 & 0.951 & 0.950 & 0.960 & 0.960 & 3.88 & 4.79 & 5.02 & 5.07 & 5.08\\ 
  SPI & 0.948 & 0.948 & 0.940 & 0.950 & 0.946 & 3.37 & 4.11 & 4.32 & 4.38 & 4.40 \\ [3pt]
    $T = 100$        &       &       &       &       &  \\[3pt]
  QPI-f & 0.901 & 0.907 & 0.912 & 0.909 & 0.906 &  3.28 & 3.85 & 3.97 & 4.01 & 4.01 \\ 
  QPI-p & 0.933 & 0.931 & 0.938 & 0.933 & 0.938 & 3.82 & 4.41 & 4.55 & 4.58 & 4.59   \\ 
  QPI-f-u & 0.901 & 0.923 & 0.931 & 0.929 & 0.932 & 3.28 & 4.07 & 4.29 & 4.35 & 4.37 \\ 
  QPI-p-u & 0.931 & 0.943 & 0.950 & 0.950 & 0.947 & 3.82 & 4.64 & 4.85 & 4.90 & 4.93  \\ 
  $L_2$-PPI-f-opv  & 0.915 & 0.925 & 0.935 & 0.936 & 0.935 & 3.52 & 4.25 & 4.43 & 4.48 & 4.50  \\ 
  $L_2$-PPI-p-opv & 0.941 & 0.948 & 0.954 & 0.955 & 0.954 & 4.17 & 4.90 & 5.07 & 5.11 & 5.13 \\ 
  $L_1$-PPI-f-opv & 0.916 & 0.926 & 0.935 & 0.936 & 0.936 & 3.53 & 4.25 & 4.43 & 4.48 & 4.50 \\ 
  $L_1$-PPI-p-opv & 0.941 & 0.947 & 0.954 & 0.952 & 0.955 & 4.17 & 4.90 & 5.07 & 5.12 & 5.13  \\ 
  SPI &  0.951 & 0.947 & 0.947 & 0.946 & 0.942 & 3.41 & 4.13 & 4.33 & 4.39 & 4.40   \\  [3pt]
    $T = 50$        &       &       &       &       &  \\[3pt]
QPI-f & 0.844 & 0.874 & 0.884 & 0.883 & 0.888 & 3.09 & 3.68 & 3.83 & 3.87 & 3.89 \\ 
  QPI-p & 0.903 & 0.921 & 0.929 & 0.929 & 0.934 & 4.01 & 4.74 & 4.85 & 4.93 & 4.95 \\ 
  QPI-f-u & 0.845 & 0.892 & 0.907 & 0.910 & 0.910 & 3.09 & 3.93 & 4.15 & 4.23 & 4.26 \\ 
  QPI-p-u & 0.905 & 0.929 & 0.934 & 0.940 & 0.946 & 4.03 & 4.91 & 5.17 & 5.23 & 5.24 \\ 
  $L_2$-PPI-f-opv & 0.871 & 0.905 & 0.917 & 0.918 & 0.922 & 3.45 & 4.19 & 4.38 & 4.46 & 4.47 \\ 
  $L_2$-PPI-p-opv & 0.934 & 0.941 & 0.948 & 0.950 & 0.954 & 4.71 & 5.48 & 5.60 & 5.67 & 5.68\\ 
  $L_1$-PPI-f-opv & 0.873 & 0.907 & 0.920 & 0.919 & 0.923 & 3.46 & 4.20 & 4.40 & 4.47 & 4.48\\ 
  $L_1$-PPI-p-opv & 0.934 & 0.942 & 0.948 & 0.950 & 0.954 & 4.69 & 5.44 & 5.57 & 5.64 & 5.64 \\ 
  SPI & 0.942 & 0.946 & 0.948 & 0.939 & 0.950 & 3.39 & 4.11 & 4.33 & 4.38 & 4.40 \\ 
        \bottomrule
    \end{tabular}\\
      \raggedright
     \textit{Note:} All PPIs with ``-opv'' symbol are based on applying under-smoothing and optimal bandwidths to estimate mean and variance functions, respectively.
    
\end{table}

\FloatBarrier
\section{Conclusions}\label{Sec:conclusion}
In this paper, we propose some forward bootstrap prediction algorithms based on the local constant estimation of the model. With theoretical and practical validations, we show our bootstrap-based point predictions work well and its MSPEs are very close to the oracle predictions. By debiasing the non-parametric estimation with under-smoothing bandwidth, we show the confidence bound for the multi-step ahead estimator can be approximated by the bootstrap. As a result, we can get a pertinence prediction interval under a specifically designed algorithm. Empirically, we further take the predictive residuals to make predictions that can alleviate the under-coverage of PI for a small sample size. Among different bootstrap-based PIs, revealed by simulation studies, the PPI with predictive residuals is the best one, which is competitive with the oracle PI.

\newpage
\newpage
\bibliographystyle{apalike}
\bibliography{refs}

\begin{thebibliography}{}

\bibitem[Bradley, 2005]{bradley2005basic}
Bradley, R.~C. (2005).
\newblock Basic properties of strong mixing conditions. a survey and some open
  questions.
\newblock {\em Probability surveys}, 2:107--144.

\bibitem[Chen et~al., 2004]{chen2004nonparametric}
Chen, R., Yang, L., and Hafner, C. (2004).
\newblock Nonparametric multistep-ahead prediction in time series analysis.
\newblock {\em Journal of the Royal Statistical Society: Series B (Statistical
  Methodology)}, 66(3):669--686.

\bibitem[Franke et~al., 2002a]{franke2002bootstrap}
Franke, J., Kreiss, J.-P., and Mammen, E. (2002a).
\newblock Bootstrap of kernel smoothing in nonlinear time series.
\newblock {\em Bernoulli}, 8(1):1--37.

\bibitem[Franke et~al., 2002b]{franke2002properties}
Franke, J., Kreiss, J.-P., Mammen, E., and Neumann, M.~H. (2002b).
\newblock Properties of the nonparametric autoregressive bootstrap.
\newblock {\em Journal of Time Series Analysis}, 23(5):555--585.

\bibitem[Franke et~al., 2004]{franke2004bootstrapping}
Franke, J., Neumann, M.~H., and Stockis, J.-P. (2004).
\newblock Bootstrapping nonparametric estimators of the volatility function.
\newblock {\em Journal of Econometrics}, 118(1-2):189--218.

\bibitem[Lee and Billings, 2003]{lee2003new}
Lee, K. and Billings, S. (2003).
\newblock A new direct approach of computing multi-step ahead predictions for
  non-linear models.
\newblock {\em International Journal of Control}, 76(8):810--822.

\bibitem[Min and Hongzhi, 1999]{min1999probabilistic}
Min, C. and Hongzhi, A. (1999).
\newblock The probabilistic properties of the nonlinear autoregressive model
  with conditional heteroskedasticity.
\newblock {\em Acta Mathematicae Applicatae Sinica}, 15(1):9--17.

\bibitem[Pan and Politis, 2016]{pan2016bootstrap}
Pan, L. and Politis, D.~N. (2016).
\newblock Bootstrap prediction intervals for linear, nonlinear and
  nonparametric autoregressions.
\newblock {\em Journal of Statistical Planning and Inference}, 177:1--27.

\bibitem[Pascual et~al., 2004]{pascual2004bootstrap}
Pascual, L., Romo, J., and Ruiz, E. (2004).
\newblock Bootstrap predictive inference for arima processes.
\newblock {\em Journal of Time Series Analysis}, 25(4):449--465.

\bibitem[Pascual et~al., 2006]{pascual2006bootstrap}
Pascual, L., Romo, J., and Ruiz, E. (2006).
\newblock Bootstrap prediction for returns and volatilities in garch models.
\newblock {\em Computational Statistics \& Data Analysis}, 50(9):2293--2312.

\bibitem[Pemberton, 1987]{pemberton1987exact}
Pemberton, J. (1987).
\newblock Exact least squares multi-step prediction from nonlinear
  autoregressive models.
\newblock {\em Journal of Time Series Analysis}, 8(4):443--448.

\bibitem[Politis, 2009]{politis2009financial}
Politis, D.~N. (2009).
\newblock Financial time series.
\newblock {\em Wiley Interdisciplinary Reviews: Computational Statistics},
  1(2):157--166.

\bibitem[Politis, 2013]{politis2013model}
Politis, D.~N. (2013).
\newblock Model-free model-fitting and predictive distributions.
\newblock {\em Test}, 22(2):183--221.

\bibitem[Politis, 2015]{politis2015model}
Politis, D.~N. (2015).
\newblock Pertinent prediction intervals.
\newblock In {\em Model-Free Prediction and Regression}, pages 43--45.
  Springer.

\bibitem[Politis, 2022]{politisstudentization}
Politis, D.~N. (2022).
\newblock Studentization vs. variance stabilization: a simple way out of an old
  dilemma.

\bibitem[Thombs and Schucany, 1990]{thombs1990bootstrap}
Thombs, L.~A. and Schucany, W.~R. (1990).
\newblock Bootstrap prediction intervals for autoregression.
\newblock {\em Journal of the American Statistical Association},
  85(410):486--492.

\bibitem[Wang and Politis, 2021]{wang2021model}
Wang, Y. and Politis, D.~N. (2021).
\newblock Model-free bootstrap and conformal prediction in regression:
  Conditionality, conjecture testing, and pertinent prediction intervals.
\newblock {\em arXiv preprint arXiv:2109.12156}.

\bibitem[Wu and Politis, 2023]{wu2023bootstrap}
Wu, K. and Politis, D.~N. (2023).
\newblock Bootstrap prediction inference of non-linear autoregressive models.
\newblock {\em arXiv preprint arXiv:2306.04126}.

\end{thebibliography}

\clearpage
\appendix

\section*{\textsc{Appendix A: Proofs}}\label{Appendix:Proof}
\begin{proof}[\textbf{\textsc{Proof of Theorem 3.1}}]
To show \cref{claim1} satisfied for $k\geq1$, we can just show the case with $k=2$. Cases with $k=1$ and $k>2$ can be handled similarly. $F_{X_{T+2}|X_T}(x)$ is equivalent to:
\begin{equation}\label{truedis}
\begin{split}
    F_{X_{T+2}|X_T}(x) &=  \mathbb{P}(X_{T+2}\leq x|X_T)\\
      &=  \mathbb{P}(m(X_{T+1}) + \sigma(X_{T+1})\epsilon_{T+2}\leq x|X_T)\\
    & = \mathbb{P}\left(\epsilon_{T+2}\leq \frac{x - m(m(X_{T})+\sigma(X_t)\epsilon_{T+1})}{\sigma(m(X_{T})+\sigma(X_t)\epsilon_{T+1})}\bigg\vert X_T\right)\\
    & = \mathbb{E}\left[\mathbb{P}\left(\epsilon_{T+2}\leq \frac{x - m(m(X_{T})+\sigma(X_{T})\epsilon_{T+1})}{\sigma(m(X_{T})+\sigma(X_{T})\epsilon_{T+1})}\bigg\vert\epsilon_{T+1},X_T\right)\bigg\vert X_T\right]\\
    & = \mathbb{E}\left[F_{\epsilon}\left(\frac{x - m(m(X_{T})+\sigma(X_{T})\epsilon_{T+1})}{\sigma(m(X_{T})+\sigma(X_{T})\epsilon_{T+1})}\right)\bigg\vert X_T\right]\\
    & = \mathbb{E}\left[F_{\epsilon}\left( \mathcal{G}(x,X_T,\epsilon_{T+1})   \right)\bigg\vert X_T\right];
\end{split}
\end{equation}
we use $\mathcal{G}(x,X_T,\epsilon_{T+1}) $ to represent $\frac{x - m(m(X_{T})+\sigma(X_{T})\epsilon_{T+1})}{\sigma(m(X_{T})+\sigma(X_{T})\epsilon_{T+1})}$ to simplify notations. Similarly, we can analyze $F_{X_{T+2}^{*}}(x)$, it has below equivalent expressions:
\begin{equation}\label{bootdis}
\begin{split}
    F_{X^*_{T+2}|X_T,\ldots,X_0}(x) &=  \mathbb{P}(X^*_{T+2}\leq x|X_T,\ldots,X_0)\\
    & = \mathbb{E}\left[\mathbb{P}\left(\hat{\epsilon}^*_{T+2}\leq \widehat{\mathcal{G}}(x, X_T,\hat{\epsilon}^*_{T+1})\bigg\vert \hat{\epsilon}^*_{T+1},X_T,\ldots,X_0\right)\bigg\vert X_T,\ldots,X_0\right]\\
    & = \mathbb{E}^*\left[\widehat{F}_{\epsilon}\left(\widehat{\mathcal{G}}(x, X_T,\hat{\epsilon}^*_{T+1})\right)\right],
\end{split}
\end{equation}
 where $\widehat{\mathcal{G}}(x, X_T,\hat{\epsilon}^*_{T+1})$ represents $\frac{x - \widehat{m}_h(\widehat{m}_h(X_T)+\widehat{\sigma}_h(X_T)\hat{\epsilon}^*_{T+1})}{\widehat{\sigma}_h(\widehat{m}_h(X_T)+\widehat{\sigma}_h(X_T)\hat{\epsilon}^*_{T+1})}$ and $\mathbb{E}^*(\cdot)$ represents the expectation in the bootstrap world, i.e., $\mathbb{E}(\cdot|X_T,\ldots,X_0)$. Thus, we hope to show:
 \begin{equation}\label{equivatheorem}
     \sup_{|x|\leq c_T}\bigg\vert \mathbb{E}^*\left[\widehat{F}_{\epsilon}(\widehat{\mathcal{G}}(x, X_T,\hat{\epsilon}^*_{T+1})) \right] - \mathbb{E}\left[F_{\epsilon}\left( \mathcal{G}(x,X_T,\epsilon_{T+1})   \right)\bigg\vert X_T\right] \bigg\vert \overset{p}{\to} 0. 
 \end{equation}
 However, it is hard to analyze the \cref{equivatheorem} since there is a random variable $X_T$ inside $\mathbb{E}^*(\cdot)$ and $\mathbb{E}(\cdot)$. Thus, we consider two regions of $X_T$, i.e., (1) $|X_T|> \gamma_T$ and (2) $|X_T|\leq\gamma_T$, where $\gamma_T$ is an appropriate sequence that converges to infinity. Under A1, A2 and A5, by Lemma 1 of \cite{franke2004bootstrapping}, we have:
 \begin{equation}\label{xTregion}
    \mathbb{P}(|X_T|> \gamma_T)\to 0.  
 \end{equation}
In addition, we have a relationship:
 \begin{equation}\label{decompose}
 \begin{split}
     &\mathbb{P}\left( \sup_{|x|\leq c_T}\bigg\vert \mathbb{E}^*\left[\widehat{F}_{\epsilon}(\widehat{\mathcal{G}}(x, X_T,\hat{\epsilon}^*_{T+1})) \right] - \mathbb{E}\left[F_{\epsilon}\left( \mathcal{G}(x,X_T,\epsilon_{T+1})   \right)\bigg\vert X_T\right] \bigg\vert > \varepsilon\right) \\
     &\leq\mathbb{P}((|X_T|> \gamma_T)) + \mathbb{P}\left((|X_T|\leq \gamma_T) \bigcap \left(\sup_{|x|\leq c_T}\bigg\vert \mathbb{E}^*\left[\widehat{F}_{\epsilon}(\widehat{\mathcal{G}}(x, X_T,\hat{\epsilon}^*_{T+1})) \right] - \mathbb{E}\left[F_{\epsilon}\left( \mathcal{G}(x,X_T,\epsilon_{T+1})   \right)\bigg\vert X_T\right] \bigg\vert>\varepsilon\right) \right).
 \end{split}
 \end{equation}
Thus, to verify \cref{equivatheorem}, we just need to show that the second term of the r.h.s. of \cref{decompose} converges to 0. We can take the sequence $c_T$ and $\gamma_T$ to be the same sequence which converges to infinity slowly enough. Then, it is enough for us to analyze the asymptotic probability of the below expression:
\begin{equation}\label{start}
    \sup_{|x|\leq c_T, |y|\leq c_T}\bigg\vert \mathbb{E}^*\left[\widehat{F}_{\epsilon}(\widehat{\mathcal{G}}(x, y,\hat{\epsilon}^*_{T+1})) \right] - \mathbb{E}\left[F_{\epsilon}\left( \mathcal{G}(x,y,\epsilon_{T+1})   \right)\right] \bigg\vert>\varepsilon.
\end{equation}
Decompose the l.h.s. of \cref{start} as:
\begin{equation}\label{intermediate}
\begin{split}
    &\sup_{|x|\leq c_T, |y|\leq c_T}\bigg\vert \mathbb{E}^*\left[\widehat{F}_{\epsilon}(\widehat{\mathcal{G}}(x, y,\hat{\epsilon}^*_{T+1})) \right] - \mathbb{E}\left[F_{\epsilon}\left( \mathcal{G}(x,y,\epsilon_{T+1})   \right)\right] \bigg\vert\\
    &= \sup_{|x|\leq c_T, |y|\leq c_T}\bigg\vert \mathbb{E}^*\left[\widehat{F}_{\epsilon}(\widehat{\mathcal{G}}(x, y,\hat{\epsilon}^*_{T+1})) \right] - \mathbb{E}^*\left[ F_{\epsilon}(\widehat{\mathcal{G}}(x, y,\hat{\epsilon}^*_{T+1})) \right] + \mathbb{E}^*\left[ F_{\epsilon}(\widehat{\mathcal{G}}(x, y,\hat{\epsilon}^*_{T+1})) \right] -  \mathbb{E}\left[F_{\epsilon}\left( \mathcal{G}(x,y,\epsilon_{T+1})   \right)\right]    \bigg\vert  \\
    & \leq \sup_{|x|\leq c_T, |y|\leq c_T}\bigg\vert \mathbb{E}^*\left[\widehat{F}_{\epsilon}(\widehat{\mathcal{G}}(x, y,\hat{\epsilon}^*_{T+1})) \right] - \mathbb{E}^*\left[ F_{\epsilon}(\widehat{\mathcal{G}}(x, y,\hat{\epsilon}^*_{T+1})) \right]   \bigg\vert \\
    &+ \sup_{|x|\leq c_T, |y|\leq c_T}\bigg\vert  \mathbb{E}^*\left[ F_{\epsilon}(\widehat{\mathcal{G}}(x, y,\hat{\epsilon}^*_{T+1})) \right] -  \mathbb{E}\left[F_{\epsilon}\left( \mathcal{G}(x,y,\epsilon_{T+1})   \right)\right]   \bigg\vert.
\end{split}
\end{equation}
Then, we analyze two terms on the r.h.s. of \cref{intermediate} separately. For the first term, we have:
\begin{equation}\label{part1}
\begin{split}
    &\sup_{|x|\leq c_T, |y|\leq c_T}\bigg\vert \mathbb{E}^*\left[\widehat{F}_{\epsilon}(\widehat{\mathcal{G}}(x, y,\hat{\epsilon}^*_{T+1})) \right] - \mathbb{E}^*\left[ F_{\epsilon}(\widehat{\mathcal{G}}(x, y,\hat{\epsilon}^*_{T+1})) \right]   \bigg\vert\\
    & \leq \sup_{|x|\leq c_T, |y|\leq c_T}\mathbb{E}^*\bigg\vert \widehat{F}_{\epsilon}(\widehat{\mathcal{G}}(x, y,\hat{\epsilon}^*_{T+1})) -  F_{\epsilon}(\widehat{\mathcal{G}}(x, y,\hat{\epsilon}^*_{T+1}))  \bigg\vert\\
    & \leq \sup_{|x|\leq c_T, |y|\leq c_T, z} \bigg\vert \widehat{F}_{\epsilon}(\widehat{\mathcal{G}}(x, y,z)) -  F_{\epsilon}(\widehat{\mathcal{G}}(x, y,z))  \bigg\vert \overset{p}{\to} 0,~\text{under \cref{resuniform}}.
\end{split}
\end{equation}
For the second term on the r.h.s. of \cref{intermediate}, we have:
\begin{equation}\label{part2}
\begin{split}
    &\sup_{|x|\leq c_T, |y|\leq c_T}\bigg\vert  \mathbb{E}^*\left[ F_{\epsilon}(\widehat{\mathcal{G}}(x, y,\hat{\epsilon}^*_{T+1})) \right] -  \mathbb{E}\left[F_{\epsilon}\left( \mathcal{G}(x,y,\epsilon_{T+1})   \right)\right]   \bigg\vert \\
    & = \sup_{|x|\leq c_T, |y|\leq c_T}\bigg\vert \frac{1}{T}\sum_{i=1}^{T}F_{\epsilon}(\widehat{\mathcal{G}}(x,y,\hat{\epsilon}_i)) -  \frac{1}{T}\sum_{i=1}^{T}F_{\epsilon}(\mathcal{G}(x,y,\epsilon_i))  + \frac{1}{T}\sum_{i=1}^{T}F_{\epsilon}(\mathcal{G}(x,y,\epsilon_i)) - \mathbb{E}\left[F_{\epsilon}\left( \mathcal{G}(x,y,\epsilon_{T+1})   \right)\right]     \bigg\vert \\
    &\leq \sup_{|x|\leq c_T, |y|\leq c_T}\bigg\vert \frac{1}{T}\sum_{i=1}^{T}F_{\epsilon}(\widehat{\mathcal{G}}(x,y,\hat{\epsilon}_i)) -  \frac{1}{T}\sum_{i=1}^{T}F_{\epsilon}(\mathcal{G}(x,y,\epsilon_i))    \bigg\vert\\
    &+ \sup_{|x|\leq c_T, |y|\leq c_T}\bigg\vert \frac{1}{T}\sum_{i=1}^{T}F_{\epsilon}(\mathcal{G}(x,y,\epsilon_i)) - \mathbb{E}\left[F_{\epsilon}\left( \mathcal{G}(x,y,\epsilon_{T+1})   \right)\right]      \bigg\vert, 
\end{split}
\end{equation}
where, $\{\epsilon_i\}_{i=1}^{T}$ are taken as $(X_i - m(X_{i-1}))/\sigma(X_{i-1})$ for $i=1,\ldots,T$. And $\{\hat{\epsilon}_i\}_{i=1}^{T}$ are computed by $(X_i - \widehat{m}(X_{i-1}))/\widehat{\sigma}(X_{i-1})$ for $i=1,\ldots,T$. We can show:
\begin{equation}\label{epandhatep}
\begin{split}
    &\mathbb{P}\left(\max_{i=1,\ldots,T}\bigg\vert\epsilon_i - \hat{\epsilon}_i\bigg\vert>\varepsilon\right)\\
    & = \mathbb{P}\left(\max_{i=1,\ldots,T}\bigg\vert \frac{X_i - m(X_{i-1})}{\sigma(X_{i-1})} - \frac{X_i - \widehat{m}(X_{i-1})}{\widehat{\sigma}(X_{i-1})}          \bigg\vert>\varepsilon\right)\\
    & \leq \mathbb{P}\left(\left(\max_{i=1,\ldots,T}|X_i|>c_T \right) \bigcup \left(\max_{i=1,\ldots,T}|X_{i-1}|>c_T\right) \right) \\
    &+ \mathbb{P}\left( \left(\max_{i=1,\ldots,T}|X_i|<c_T\right) \bigcap \left(\max_{i=1,\ldots,T}|X_{i-1}|<c_T\right) \bigcap \left(\max_{i=1,\ldots,T}\bigg\vert \frac{X_i - m(X_{i-1})}{\sigma(X_{i-1})} - \frac{X_i - \widehat{m}(X_{i-1})}{\widehat{\sigma}(X_{i-1})}        \bigg\vert>\varepsilon\right)\right) \\
    & \leq o(1) + \mathbb{P}\left(\sup_{|x|,|y| \leq c_T}\bigg\vert \frac{x - m(y)}{\sigma(y)} - \frac{x - \widehat{m}(y)}{\widehat{\sigma}(y)}     \bigg\vert >\varepsilon\right) \\
    & \to 0.
\end{split}
\end{equation}
We further consider two terms on the r.h.s. of \cref{part2} separately. For the first term, by Taylor expansion, we have:
\begin{equation}\label{part2part1}
\begin{split}
    &\sup_{|x|\leq c_T, |y|\leq c_T}\bigg\vert \frac{1}{T}\sum_{i=1}^{T}F_{\epsilon}(\widehat{\mathcal{G}}(x,y,\hat{\epsilon}_i)) -  \frac{1}{T}\sum_{i=1}^{T}F_{\epsilon}(\mathcal{G}(x,y,\epsilon_i))    \bigg\vert\\
    & = \sup_{|x|\leq c_T, |y|\leq c_T}\bigg\vert \frac{1}{T}\sum_{i=1}^{T}\left( F_{\epsilon}(\mathcal{G}(x,y,\epsilon_i)) + f_{\epsilon}(o_i)(\widehat{\mathcal{G}}(x,y,\hat{\epsilon}_i) - \mathcal{G}(x,y,\epsilon_i))  \right)  -  \frac{1}{T}\sum_{i=1}^{T}F_{\epsilon}(\mathcal{G}(x,y,\epsilon_i)) \bigg\vert \\
    & = \sup_{|x|\leq c_T, |y|\leq c_T}\bigg\vert \frac{1}{T}\sum_{i=1}^{T} f_{\epsilon}(o_i)(\widehat{\mathcal{G}}(x,y,\hat{\epsilon}_i) - \mathcal{G}(x,y,\epsilon_i))     \bigg\vert \\
    & \leq \sup_{|x|\leq c_T, |y|\leq c_T} \frac{1}{T}\sum_{i=1}^{T}\bigg\vert  f_{\epsilon}(o_i)(\widehat{\mathcal{G}}(x,y,\hat{\epsilon}_i) - \mathcal{G}(x,y,\epsilon_i))\bigg\vert  \\ 
    &\leq \sup_{|x|\leq c_T, |y|\leq c_T} \sup_{z}|f_{\epsilon}(z)|\cdot\frac{1}{T}\sum_{i=1}^{T}\bigg\vert  \widehat{\mathcal{G}}(x,y,\hat{\epsilon}_i) - \mathcal{G}(x,y,\epsilon_i)\bigg\vert\\
    & \leq  \sup_{|x|\leq c_T, |y|\leq c_T} C\cdot\frac{1}{T}\sum_{i=1}^{T}\bigg\vert  \widehat{\mathcal{G}}(x,y,\hat{\epsilon}_i) - \mathcal{G}(x,y,\epsilon_i)    \bigg\vert~\text{(under A7)}\\
    & \leq \sup_{|x|\leq c_T, |y|\leq c_T,j\in\{1,\ldots,T\}} C\cdot\bigg\vert  \widehat{\mathcal{G}}(x,y,\hat{\epsilon}_j) - \mathcal{G}(x,y,\epsilon_j)   \bigg\vert.
\end{split} 
\end{equation}
 From \cref{epandhatep} and \cref{Lemma:uniformest}, we have \cref{part2part1} converges to 0 in probability. For the second term on the r.h.s. of \cref{part2}, by the uniform law of large numbers, we have: 
 \begin{equation}
     \sup_{|x|\leq c_T, |y|\leq c_T}\bigg\vert \frac{1}{T}\sum_{i=1}^{T}F_{\epsilon}(\mathcal{G}(x,y,\epsilon_i)) - \mathbb{E}\left[F_{\epsilon}\left( \mathcal{G}(x,y,\epsilon_{T+1})   \right)\right]      \bigg\vert \overset{p}{\to} 0. 
 \end{equation}
 Combine all pieces, \cref{start} converges to 0 in probability, which implies \cref{Theorem:QPI}. 
 \end{proof}
\begin{proof}[\textbf{\textsc{Proof of Theorem 3.2}}]
we want to show:
\begin{equation}
    \sup_{|x|\leq c_{T}}\left| F_{X^*_{T+k}- \widehat{X}^*_{T+k}|X_T,\ldots,X_0}(x) -  F_{X_{T+k} - \widehat{X}_{T+k}|X_T,\ldots,X_0}(x)\right| \overset{p}{\to} 0,~\text{for}~k \geq 1,
\end{equation}
where $X^*_{T+k}- \widehat{X}^*_{T+k}$ and $X_{T+k} - \widehat{X}_{T+k}$ are predictive roots. We still present the proof for the two-step prediction. Proof for higher step prediction can be shown similarly. When we are dealing with two-step ahead predictions, predictive roots have the same expression as in \cref{pretoot2stepreal,pretoot2stepboot}. Thus, we want to measure the asymptotic distance between below two quantities:
\begin{equation}\label{ptwopreroot}
\begin{split}
    &\mathbb{P}\left(m(m(X_T)+\epsilon_{T+1}) + \epsilon_{T+2} - \frac{1}{M}\sum_{j=1}^{M} \widehat{m}_{h}\left(\widehat{m}_{h}(X_T)+\hat{\epsilon}_{j,T+1}\right)\leq x\bigg\vert X_{T},\ldots,X_{0}\right);\\
    &\mathbb{P}\left(\widehat{m}_{h}(\widehat{m}_{h}(X_T)+\hat{\epsilon}^*_{T+1}) + \hat{\epsilon}^*_{T+2} - \frac{1}{M}\sum_{j=1}^{M}  \widehat{m}^*_{h}\left(\widehat{m}^*_{h}(X_T)+\hat{\epsilon}^*_{j,T+1}\right)\leq x\bigg\vert X_{T},\ldots,X_{0}\right).
\end{split}
\end{equation}
Compare to \cref{truedis,bootdis}, \cref{pretoot2stepreal,pretoot2stepboot} just have two more terms $\frac{1}{M}\sum_{j=1}^{M} \widehat{m}_{h}\left(\widehat{m}_{h}(X_T)+\hat{\epsilon}_{j,T+1}\right)$ and $\frac{1}{M}\sum_{j=1}^{M}  \widehat{m}^*_{h}\left(\widehat{m}^*_{h}(X_T)+\hat{\epsilon}^*_{j,T+1}\right)$ in predictive root in the real and bootstrap world, respectively. By the LLN, these two terms converge to their corresponding mean in the real or bootstrap world. Based on the consistency between $\widehat{m}_{h}(\cdot)$ and $\widehat{m}^*_{h}(\cdot)$, we can show \cref{Theorem:PPI_part1} similarly with the procedure in proving \cref{Theorem:QPI}. 
 \end{proof}
 
\begin{proof}[\textbf{\textsc{Proof of Theorem 3.3}}]
The proof is based on $\{X_0,\ldots,X_T\}\in \Omega_T$. We need to verify \cref{kstepestdist}, i.e., we can build confidence bound for the $k$-step ahead estimation by bootstrap. Still, we focus on two-step ahead prediction, i.e., we want to show:
\begin{equation}\label{2stepestdist}
\begin{split}
    \sup_{|x|\leq c_{T}} & \left| \mathbb{P}\left( \sqrt{Th}\left(\widehat{m}_{h}(\widehat{m}_{h}(X_T)) -  m(m(X_T) \right)\leq x  \right) -  \right. \\
    &\left. \mathbb{P}\left( \sqrt{Th}\left( \widehat{m}^*_{h}(\widehat{m}^*_{h}(X_T))- \widehat{m}_{h}(\widehat{m}_{h}(X_T))   \right)\leq x  \right)
     \right| \overset{p}{\to} 0.
\end{split}
\end{equation}
Apply the property $\mathbb{P}(|X_T|> c_T)\to 0$ again, it is enough to show:
\begin{equation}\label{2stepestdistsim}
\begin{split}
    \sup_{|x|, |y| \leq c_{T}} & \left| \mathbb{P}\left( \sqrt{Th}\left(\widehat{m}_{h}(\widehat{m}_{h}(y)) -  m(m(y))   \right)\leq x  \right) -  \right. \\
    &\left. \mathbb{P}\left( \sqrt{Th}\left( \widehat{m}^*_{h}(\widehat{m}^*_{h}(y))- \widehat{m}_{h}(\widehat{m}_{h}(y))   \right)\leq x  \right)
     \right| \overset{p}{\to} 0.
\end{split}
\end{equation}
 To handle the uniform convergence on $y$, we make a $\varepsilon$-covering of $X_T$. Let the $\varepsilon$-covering number of $[-c_T,c_T]$ be $C_{N} = N(\varepsilon;[-c_T,c_T];|\cdot|)$ which means for every $y\in[-c_T,c_T]$, $\exists~i\in\{1,2,\ldots, C_{N}\}$ s.t. $|y - y^{i}|\leq \varepsilon$ for $\forall \varepsilon>0$. Define $y_0 \in \{y^{1},\ldots,y^{C_{N}}\}$, we can consider:
\begin{equation}\label{2stepestdistsp}
\begin{split}
    &\sup_{|x|, |y| \leq c_{T}}  \left| \mathbb{P}\left( \sqrt{Th}\left(\widehat{m}_{h}(\widehat{m}_{h}(y)) -  m(m(y)  \right)\leq x  \right) - \right. \\
    &\left. \mathbb{P}\left( \sqrt{Th}\left( \widehat{m}^*_{h}(\widehat{m}^*_{h}(y))- \widehat{m}_{h}(\widehat{m}_{h}(y))   \right)\leq x  \right)
     \right| \leq \\
    &\sup_{|x|, |y| \leq c_{T}}  \left|\mathbb{P}\left( \sqrt{Th}\left(\widehat{m}_{h}(\widehat{m}_{h}(y)) -  m(m(y))   \right)\leq x  \right)  - \mathbb{P}\left( \sqrt{Th}\left(\widehat{m}_{h}(\widehat{m}_{h}(y_0)) -  m(m(y_0))   \right)\leq x  \right) \right|\\
    &  + \sup_{\substack{|x| \leq c_{T},\\ y_0 \in \{y^{1},\ldots,y^{C_{N}}\}}}  \left|  \mathbb{P}\left( \sqrt{Th}\left(\widehat{m}_{h}(\widehat{m}_{h}(y_0)) -  m(m(y_0))   \right)\leq x  \right) \right.  \\
    & \left.- \mathbb{P}\left( \sqrt{Th}\left( \widehat{m}^*_{h}(\widehat{m}^*_{h}(y_0))- \widehat{m}_{h}(\widehat{m}_{h}(y_0))   \right)\leq x  \right)      \right|     \\
    & + \sup_{|x|, |y| \leq c_{T}}  \left|\mathbb{P}\left( \sqrt{Th}\left( \widehat{m}^*_{h}(\widehat{m}^*_{h}(y_0))- \widehat{m}_{h}(\widehat{m}_{h}(y_0))   \right)\leq x  \right)  - \mathbb{P}\left( \sqrt{Th}\left( \widehat{m}^*_{h}(\widehat{m}^*_{h}(y))- \widehat{m}_{h}(\widehat{m}_{h}(y))   \right)\leq x  \right)   \right|.
\end{split}
\end{equation}
For the first term on the r.h.s. of \cref{2stepestdistsp}, we have:
\begin{equation}\label{rootpart}
\begin{split}
    & \sup_{|x|, |y| \leq c_{T}}  \left|\mathbb{P}\left( \sqrt{Th}\left(\widehat{m}_{h}(\widehat{m}_{h}(y)) -  m(m(y))   \right)\leq x  \right)  - \mathbb{P}\left( \sqrt{Th}\left(\widehat{m}_{h}(\widehat{m}_{h}(y_0)) -  m(m(y_0))   \right)\leq x  \right) \right|\\
    &=\sup_{|x|, |y| \leq c_{T}}  \left|\mathbb{P}\left( \sqrt{Th}\left(\widehat{m}_{h}(\widehat{m}_{h}(y_0)) -  m(m(y_0))   \right) + \sqrt{Th}\left(C_1(\widehat{m}_h(y) - \widehat{m}_h(y_0)) + C_2(m(y_0) - m(y)) \right)\leq x  \right) \right. \\
    &\left.- \mathbb{P}\left( \sqrt{Th}\left(\widehat{m}_{h}(\widehat{m}_{h}(y_0)) -  m(m(y_0))   \right)\leq x  \right) \right|.\\
\end{split}
\end{equation}
where, $C_1$ and $C_2$ are some finite constant since the derivative of $\widehat{m}_h(\cdot)$ and $m(\cdot)$ are bounded. 
Consider the first term inside the absolute bracket of the r.h.s. \cref{rootpart}, we can think this is a convolution of two random variables:
\begin{equation}
\begin{split}
    &\mathbb{P}\left( \sqrt{Th}\left(\widehat{m}_{h}(\widehat{m}_{h}(y_0)) -  m(m(y_0))   \right) + \sqrt{Th}\left(C_1(\widehat{m}_h(y) - \widehat{m}_h(y_0)) + C_2(m(y_0) - m(y))\right)\leq x  \right)\\
    & = \mathbb{P}\left( X + Z \leq x    \right).
\end{split}
\end{equation}
Further, based on the smoothing property of $\widehat{m}_h(\cdot)$ and $m(\cdot)$ again, we can take $\varepsilon$ small enough to make random variable $Z$ close to be degenerated, i.e., $\mathbb{P}(Z = 0) = 1 -\mathbb{P}(Z\in A) = 1 - o(1)$; $A$ is a small set around 0 without containing 0. Thus \cref{rootpart} can be written as:
\begin{equation}
\begin{split}
    &\sup_{|x|, |y| \leq c_{T}}\left| \mathbb{P}\left( X + Z \leq x    \right) - \mathbb{P}(X\leq x)      \right| \\
    &= \sup_{|x|, |y| \leq c_{T}}\left| \mathbb{P}(X + 0 \leq x, Z = 0) + \mathbb{P}(X + Z \leq x, Z \in A) -   \mathbb{P}(X\leq x)   \right|\\
    & \leq \sup_{|x|, |y| \leq c_{T}}\left| \mathbb{P}(X\leq x) + o(1) + o(1) -   \mathbb{P}(X\leq x)   \right|\\
    & = o(1).
\end{split}
\end{equation}

Similarly, the last term on the r.h.s. of \cref{2stepestdistsp} can also be made to converge to 0. We can then focus on analyzing the middle term. In other words, it is enough to analyze the pointwise convergence property between distribution in the real and bootstrap worlds. According to the idea in estimating the distribution of non-parametric estimation by the bootstrap in the work of \cite{franke2002bootstrap}, we decompose $\sqrt{Th}\left(\widehat{m}_{h}(\widehat{m}_{h}(y_0)) -  m(m(y_0))   \right)$ as bias-type and variance terms:
\begin{equation}
\begin{split}
    &\sqrt{Th}\left(\widehat{m}_{h}(\widehat{m}_{h}(y_0)) -  m(m(y_0))   \right) \\
    & = \sqrt{Th}\left( \frac{\sum_{t=0}^{T-1}K_h(\widehat{m}_h(y_0)-X_t)X_{t+1}}{T\hat{f}_{h}(\widehat{m}_{h}(y_0))}  -   \frac{\sum_{t=0}^{T-1}K_h(\widehat{m}_h(y_0)-X_t)\cdot m(m(y_0)) }{T\hat{f}_{h}(\widehat{m}_{h}(y_0))}     \right)\\
    &= \sqrt{Th}\left( \frac{\hat{r}_{V,h}(\widehat{m}_{h}(y_0))}{\hat{f}_{h}(\widehat{m}_{h}(y_0))} +  \frac{\hat{r}_{B,h}(\widehat{m}_{h}(y_0))}{\hat{f}_{h}(\widehat{m}_{h}(y_0))} \right),
\end{split}
\end{equation}
where 
\begin{equation}
\begin{split}
    &\hat{r}_{V,h}(\widehat{m}_{h}(y_0)) = \frac{1}{T}\sum_{t=0}^{T-1}K_h(\widehat{m}_h(y_0)-X_t)\epsilon_{t+1};\\
    &\hat{r}_{B,h}(\widehat{m}_{h}(y_0)) =  \frac{1}{T}\sum_{t=0}^{T-1}K_h(\widehat{m}_h(y_0)-X_t)\left(m(X_t) -  m(m(y_0))    \right),
\end{split}
\end{equation}
where $K_h(\cdot)$ represents the function in form $\frac{1}{h}K(\cdot/h)$. Do it also for $\sqrt{Th}\left( \widehat{m}^*_{h}(\widehat{m}^*_{h}(y_0))- \widehat{m}_{h}(\widehat{m}_{h}(y_0))   \right)$, we can get:
\begin{equation}
\sqrt{Th}\left( \widehat{m}^*_{h}(\widehat{m}^*_{h}(y_0))- \widehat{m}_{h}(\widehat{m}_{h}(y_0))   \right) = \sqrt{Th}\left( \frac{\hat{r}^*_{V,h}(\widehat{m}^*_{h}(y_0))}{\hat{f}^*_{h}(\widehat{m}^*_{h}(y_0))} +  \frac{\hat{r}^*_{B,h}(\widehat{m}^*_{h}(y_0))}{\hat{f}^*_{h}(\widehat{m}^*_{h}(y_0))} \right),
\end{equation}
where 
\begin{equation}
\begin{split}
    &\hat{r}^*_{V,h}(\widehat{m}^*_{h}(y_0)) = \frac{1}{T}\sum_{t=0}^{T-1}K_h(\widehat{m}^*_h(y_0)-X^*_t)\hat{\epsilon}^*_{t+1};\\
    &\hat{r}^*_{B,h}(\widehat{m}^*_{h}(y_0)) =  \frac{1}{T}\sum_{t=0}^{T-1}K_h(\widehat{m}^*_h(y_0)-X^*_t)\left(\widehat{m}_h(X^*_t) -  \widehat{m}_h(\widehat{m}_h(y_0))    \right).
\end{split}
\end{equation}
For the variance term, by the Lemma 4.4 of \cite{franke2002bootstrap}, we have:
\begin{equation}\label{varpart}
\begin{split}
    &\sup_{x}\left|\mathbb{P}(\sqrt{Th}\hat{r}_{V,h}(x_0) \leq x) - \mathbb{P}(Z(x_0)\leq x)     \right| = o(1);\\
    & \sup_{x}\left|\mathbb{P}(\sqrt{Th}\hat{r}^*_{V,h}(x_0) \leq x) - \mathbb{P}(Z(x_0)\leq x)     \right| = op(1),
\end{split}
\end{equation}
where $Z(x_0)$ has distribution $N(0,\tau^2(x_0))$; $\tau^2(x_0) = f_{X}(x_0)\int K^2(v)dv$; $x_0\in \mathbb{R}$. Since $\widehat{m}_{h}(y_0)$ and $\widehat{m}^*_{h}(y_0)$ all converge to $m(y_0)$ in probability and the target distribution is continuous, by continuous mapping theorem, we can get the uniform convergence between the distribution of $\sqrt{Th}\hat{r}_{V,h}(m(y_0))$ and $\sqrt{Th}\hat{r}_{V,h}(\widehat{m}_{h}(y_0))$, i.e.:
\begin{equation}\label{unifconvarpart1}
    \sup_{x}\left|\mathbb{P}(\sqrt{Th}\hat{r}_{V,h}(\widehat{m}_h(y_0)) \leq x) - \mathbb{P}(\sqrt{Th}\hat{r}_{V,h}(m(y_0))\leq x)     \right| = o(1).
\end{equation}
To show the uniform convergence relationship between $\sqrt{Th}\hat{r}^*_{V,h}(m(y_0))$ and $\sqrt{Th}\hat{r}^*_{V,h}(\widehat{m}^*_{h}(y_0))$, we need the continuous $\sqrt{Th}\hat{r}^*_{V,h}(m(y_0))$, which is a convolution of $i.i.d.$ random variables $\{\hat{\epsilon}^*_{i}\}_{i = 1}^{T}\sim \widehat{F}_{\epsilon}$. Unfortunately, $\widehat{F}_{\epsilon}$ is the empirical distribution of residuals which is discrete. For making analysis more convenient, we take a convolution approach to smooth the distribution of empirical residuals, i.e., we define another random variable which is the sum of $\hat{\epsilon}$ and a standard normal random variable $\xi$:
\begin{equation}
    \Tilde{\epsilon} = \hat{\epsilon} + \xi,
\end{equation}
where $\xi\sim N(0,\mathcal{L}(T))$ and $\mathcal{L}(T)\overset{}{\to}0$ in an appropriate rate. It is easy to show that the distribution of $\Tilde{\epsilon}$, $\widetilde{F}_{\epsilon}$ is asymptotically equivalent to $\widehat{F}_{\epsilon}$, i.e., \cref{resuniform} is also satisfied for $\widetilde{F}_{\epsilon}$. In practice, we can take $\mathcal{L}(T)$ to be small enough and then we still bootstrap time series based on $\widehat{F}_{\epsilon}$ in practice. However, from a theoretical view, we would like to take $\widetilde{F}_{\epsilon}$. For simplifying the notation, we use $\widehat{F}_{\epsilon}$ throughout this paper and its representation will change according to the context.

Combine all pieces, we can get:
\begin{equation}
\begin{split}
    &\sup_{x}\left|\mathbb{P}(\sqrt{Th}\hat{r}_{V,h}(\widehat{m}_{h}(y_0)) \leq x) - \mathbb{P}(Z(m(y_0))\leq x)     \right| = op(1);\\
    & \sup_{x}\left|\mathbb{P}(\sqrt{Th}\hat{r}^*_{V,h}(\widehat{m}^*_{h}(y_0)) \leq x) - \mathbb{P}(Z(m(y_0))\leq x)     \right| = op(1).
\end{split}
\end{equation}

Then, it is left to analyze the bias-type term in the real and bootstrap worlds. We first consider the bias-type term $\hat{r}_{B,h}(\widehat{m}_{h}(y_0))$:
\begin{equation}\label{biasreal1}
\begin{split}
    &\sqrt{Th}\hat{r}_{B,h}(\widehat{m}_h(y_0)) \\
    &= \sqrt{\frac{h}{T}}\sum_{t=0}^{T-1}K_h(\widehat{m}_h(y_0)-X_t)\cdot\left(m(X_t) -  m(m(y_0))    \right)\\
    &= \sqrt{\frac{h}{T}}\sum_{t=0}^{T-1}\left[ K_h(m(y_0)- X_t) +K^{(1)}_h(\Hat{x})\cdot (\widehat{m}_{h}(y_0) - m(y_0)) \right]\cdot\left(m(X_t) -  m(m(y_0))    \right)\\
    & =   \sqrt{\frac{h}{T}}\sum_{t=0}^{T-1}K_h(m(y_0)- X_t)\cdot\left(m(X_t) -  m(m(y_0))    \right)    \\
    & + \sqrt{\frac{h}{T}}\sum_{t=0}^{T-1}K^{(1)}_h(\Hat{x})\cdot (\widehat{m}_{h}(y_0) - m(y_0)) \cdot\left(m(X_t) -  m(m(y_0))    \right).   \\
\end{split}
\end{equation}
For the first term on the r.h.s. of \cref{biasreal1}, by the ergodicity of $\{X_t\}$ series, we can find the mean of this term is:
\begin{equation}\label{biaspart1}
\begin{split}
    &\mathbb{E}\left[\sqrt{\frac{h}{T}}\sum_{t=0}^{T-1}K_h(m(y_0)- X_t)\cdot\left(m(X_t) -  m(m(y_0))    \right) \right]\\
    &=\mathbb{E}\left[\sqrt{Th}\mathbb{E}\left[ K_h(m(y_0)- X_1)\cdot\left(m(X_1) -  m(m(y_0))    \right) |X_0    \right] \right]\\
    & = \mathbb{E}\left[\sqrt{Th}\int K(v) \cdot\left(m(vh + m(y_0)) -  m(m(y_0))    \right) \cdot f_{\epsilon}(vh + m(y_0) - m(X_0)) dv\right]  \\
    & = \mathbb{E}\left[\sqrt{Th}\int K(v) \cdot\left(m^{(1)}(m(y_0))vh + m^{(2)}(\Hat{y})\cdot v^2h^2  \right) \cdot \left( f_{\epsilon}(m(y_0)-m(X_0)) + f^{(1)}_{\epsilon}(\Hat{x})\cdot vh  \right) dv \right].
\end{split}  
\end{equation}
If we take bandwidth satisfying $Th^5 \to 0$, \cref{biaspart1} converges to 0. Then, we consider the mean of the second term on the r.h.s. of \cref{biasreal1}:
\begin{equation} \label{biaspart2}
\begin{split}
    & \mathbb{E}\left[\sqrt{\frac{h}{T}}\sum_{t=0}^{T-1}K^{(1)}_h(\Hat{x})\cdot (\widehat{m}_{h}(y_0) - m(y_0)) \cdot\left(m(X_t) -  m(m(y_0))    \right)  \right] \\
    &= \sqrt{\frac{h}{T}}\sum_{t=0}^{T-1}\mathbb{E}\left[ K^{(1)}_h(\Hat{x})\cdot (\widehat{m}_{h}(y_0) - m(y_0)) \cdot\left(m(X_t) -  m(m(y_0))    \right)  \right] \\
    & = \frac{1}{T}\sum_{t=0}^{T-1}\mathbb{E}\left[\mathbb{E}\left[\sqrt{Th}\cdot K^{(1)}_h(\Hat{x})\cdot (\widehat{m}_{h}(y_0) - m(y_0)) \cdot\left(m(X_t) -  m(m(y_0))    \right) \bigg\vert X_t  \right]\right].
\end{split}
\end{equation}
Since $\mathbb{E}(\sqrt{Th}\cdot (\widehat{m}_{h}(y_0) -  m(y_0) )$ is $O(\sqrt{Th^5})$; see Lemma 4.6 of \cite{franke2002bootstrap} for a proof. Under the assumption that $K(\cdot)$ has bounded derivative and $m(\cdot)$ is bounded in a compact set, we have $\mathbb{E}(\mathbb{E}(\sqrt{Th}\cdot K^{(1)}_h(\Hat{x})\cdot (\widehat{m}_{h}(y_0) $$- m(y_0)) \cdot\left(m(X_t) -  m(m(y_0)) \right) \vert X_t ))$ is $O(\sqrt{Th^5})$; Once we select the under-smoothing bandwidth which satisfies $Th^5 \to 0$, \cref{biaspart2} converges to 0. Then, we need to analyze the variance of $\sqrt{Th}\hat{r}_{B,h}(\widehat{m}_h(y_0))$. Similarly, we can show it is $op(1)$. All in all, $\sqrt{Th}\hat{r}_{B,h}(\widehat{m}_h(y_0))$ converges to 0 in probability.

For the bias-type term $\hat{r}^*_{B,h}(\widehat{m}^*_{h}(y_0))$ in the bootstrap world, we can do a similar decomposition as we did in \cref{biasreal1}, then we can get:
\begin{equation}\label{biasboot1}
\begin{split}
    &\sqrt{Th}\hat{r}^*_{B,h}(\widehat{m}^*_h(y_0)) \\
    & =   \sqrt{\frac{h}{T}}\sum_{t=0}^{T-1}K_h(\widehat{m}_{h}(y_0)- X^*_t)\cdot\left(\widehat{m}_h(X^*_t) -  \widehat{m}_h(\widehat{m}_h(y_0))    \right)    \\
    & + \sqrt{\frac{h}{T}}\sum_{t=0}^{T-1}K^{(1)}_h(\Hat{x})\cdot (\widehat{m}^*_{h}(y_0) - \widehat{m}_h(y_0)) \cdot\left(\widehat{m}_h(X^*_t) -  \widehat{m}_h(\widehat{m}_h(y_0))    \right).   \\
\end{split}
\end{equation}

We first rely on the fact that $\mathbb{E}^{*}(\mathbb{E}^{*}(\sqrt{Th}\cdot K^{(1)}_h(\Hat{x})\cdot (\widehat{m}^*_{h}(y_0) - \widehat{m}_h(y_0)) \cdot\left(\widehat{m}_h(X^*_t) -  \widehat{m}_h(\widehat{m}_h(y_0))\right) \vert X^*_t))$ is also $O(\sqrt{Th^5})$; see Lemma 4.6 of \cite{franke2002bootstrap} for more details. Thus, taking the under-smoothing bandwidth strategy, the second term on the r.h.s. of \cref{biasboot1} also converges to 0. For the first term, we can rely on the fact that the bootstrap series is also ergodic with high probability; see Theorem 2 of \cite{franke2002properties,franke2004bootstrapping} for time series model with homoscedastic or heteroscedastic errors, respectively. Thus, with a similar analysis of the variant in the real world, we can see the bias-type term in the bootstrap world also converges to 0 in probability. Under the consistent relationship between $\hat{f}_{h}(\widehat{m}_{h}(y_0))$ and $\hat{f}^*_{h}(\widehat{m}_{h}(y_0))$ which is implied by Lemma 4.5 of \cite{franke2002bootstrap}, \cref{2stepestdist} follows from the analysis of variance and bias-type terms in the real and bootstrap world. 

\end{proof}

\clearpage
\section*{\textsc{Appendix B: The advantage of applying under-smoothing bandwidth for QPI with finite sample}}\label{Appendix:advanQPI}
The proof of \cref{Theorem:QPI} provides a \textit{big picture} of the asymptotic validity of QPI. Although the choice of the bandwidth does not influence the asymptotic validity of QPI, we can find that the QPI with under-smoothing bandwidth has a better CVR for multi-step ahead predictions from the simulation results.
%As we know, the under-smoothing bandwidth can kill the bias-type term for building the confidence interval of non-parametric estimations. However, this bias term vanishes for QPI since what we want is the consistency between the distribution of prediction and true future values, i.e., we want to show:
We attempt to analyze this phenomenon informally. Starting from the convergence result we wanted to show:
\begin{equation}\label{claim1_appendix}
    \sup_{|x|\leq c_{T}}\left| F_{X^*_{T+k}|X_T,\ldots,X_0}(x) -  F_{X_{T+k}|X_T}(x)\right| \overset{p}{\to} 0,~\text{for}~k \geq 1.\end{equation}
We still take the case with $k = 2$ as an example. From analyses in the proof of \cref{Theorem:QPI}, we can get:
\begin{equation}\label{under1}
    \sup_{|x|\leq c_{T}}\left| F_{X^*_{T+2}|X_T,\ldots,X_0}(x) -  F_{X_{T+2}|X_T}(x)\right| \leq op(1) + \sup_{|x|\leq c_T, |y|\leq c_T,j\in\{1,\ldots,T\}} C\cdot\bigg\vert  \widehat{\mathcal{G}}(x,y,\hat{\epsilon}_j) - \mathcal{G}(x,y,\epsilon_j)   \bigg\vert.
\end{equation}
Recall that $\mathcal{G}(x,X_T,\epsilon_{T+1}) $ represent $\frac{x - m(m(X_{T})+\sigma(X_{T})\epsilon_{T+1})}{\sigma(m(X_{T})+\sigma(X_{T})\epsilon_{T+1})}$ and $\widehat{\mathcal{G}}(x, X_T,\hat{\epsilon}^*_{T+1})$ represents $\frac{x - \widehat{m}_h(\widehat{m}_h(X_T)+\widehat{\sigma}_h(X_T)\hat{\epsilon}^*_{T+1})}{\widehat{\sigma}_h(\widehat{m}_h(X_T)+\widehat{\sigma}_h(X_T)\hat{\epsilon}^*_{T+1})}$. For simplifying the notation, we consider the model when $\sigma(x) \equiv 1$. Then, \cref{under1} becomes:
\begin{equation}
\begin{split}
    &\sup_{|x|\leq c_{T}}\left| F_{X^*_{T+k}|X_T,\ldots,X_0}(x) -  F_{X_{T+k}|X_T}(x)\right| \leq op(1) \\
    &+ \sup_{|y|\leq c_T,j\in\{1,\ldots,T\}} C\cdot\bigg\vert  \widehat{m}_h(\widehat{m}_h(y)+\hat{\epsilon}^*_{j}) - m(m(y)+\epsilon_{j})\bigg\vert.
\end{split}
\end{equation}

Then we can focus on analyzing $\widehat{m}_h(\widehat{m}_h(X_T)+\hat{\epsilon}^*_{j}) - m(m(X_{T})+\epsilon_{j})$. By Taylor expansion, we can get:
\begin{equation}\label{under2}
\begin{split}
&\widehat{m}_h(\widehat{m}_h(y)+\hat{\epsilon}^*_{j}) - m(m(y)+\epsilon_{j})\\
&=\widehat{m}_h(m(y)+\epsilon_{j}) - m(m(y)+\epsilon_{j}) \\
& + \widehat{m}^{(1)}_h(\Hat{\Hat{x}})(\widehat{m}_h(y)+\hat{\epsilon}^*_{j} - m(y) -\epsilon_{j}).
\end{split}
\end{equation}
For the first term of the r.h.s. of \cref{under2}, by the ergodicity, asymptotically, we have:
\begin{equation}\label{under3}
\begin{split}
    &\widehat{m}_h(m(y)+\epsilon_{j}) = \frac{\frac{1}{Th}\sum_{i=0}^{T-1}K\left(\frac{m(y) + \epsilon_{j} - X_{i}}{h}   \right)X_{i+1}}{\widehat{f}_{h}(m(y)+\epsilon_j)}\\
    & \frac{\frac{1}{Th}\sum_{i=0}^{T-1}K\left( \frac{m(y) + \epsilon_{j} - X_{i}}{h}   \right)(m(X_i)+\epsilon_{i+1})}{\widehat{f}_{h}(m(y)+\epsilon_j)} \\
    & = \frac{1}{\widehat{f}_{h}(m(y)+\epsilon_j)}\left(\frac{1}{h}\mathbb{E}(K\left(\frac{m(y) + \epsilon_{j} - X_1}{h}\right)m(X_1)) + \frac{1}{h}\mathbb{E}(K\left( \frac{m(y) + \epsilon_{j} - X_1}{h}\right)\epsilon_1) \right)\\
    & = \frac{1}{\widehat{f}_{h}(m(y)+\epsilon_j)}\left(\frac{1}{h}\int K\left(\frac{u - m(y) - \epsilon_{j}}{h}\right)m(u)f_{X}(u)du   + 0\right), (K(\cdot)~\text{are assumed to be symmetric})\\
    & = \frac{1}{\widehat{f}_{h}(m(y)+\epsilon_j)}\left(\int K\left(v\right)m(vh + m(y)+\epsilon_j)f_{X}(vh + m(y)+\epsilon_j)dv   + 0\right)\\
    & = \frac{1}{\widehat{f}_{h}(m(y)+\epsilon_j)}\left(\int K\left(v\right)[m(m(y)+\epsilon_j)+vhm^{(1)}(m(y)+\epsilon_j)+v^2h^2m^{(2)}(\Hat{\Hat{y}})] \cdot    \right. \\
    & \left. [f_{X}(m(y)+\epsilon_j) + vhf^{(1)}_{X}(m(y)+\epsilon_j) + v^2h^2f^{(2)}_{X}(\Hat{\Hat{z}})  ]dv     \right)\\
    & = \frac{1}{\widehat{f}_{h}(m(y)+\epsilon_j)}\left(m(m(y)+\epsilon_j)f_{X}(m(y)+\epsilon_j) + O(h^2) \right).
\end{split}  
\end{equation}
 The convergence of $\widehat{f}_{h}(m(y)+\epsilon_j)$ to $f_{X}(m(y)+\epsilon_j)$ guarantee the consistency relationship of \cref{under3} and $ m(m(y)+\epsilon_{j})$. Similarly, for the third term of the r.h.s. of \cref{under2}, we can do a similar analysis to find the convergence to 0 in probability. Moreover, the convergence speed is related to $O(h^2)$. When the multiple-step ahead predictions are required, we will get more and more such $O(h^2)$ terms. If we have large enough data, it is ``safe'' to focus on bandwidth with optimal rate to estimate the model. However, for the finite sample cases, it is better to take an under-smoothing $h$, though the corresponding LEN of the Prediction interval will get larger due to the mean-variance trade-off. This conclusion coincides with results shown in \cref{Tab:model1,Tab:model2}. From there, we can observe that the one-step ahead QPI with optimal bandwidth has better CVR compared to the version with under-smoothing bandwidth. Meantime, the LEN of PI with optimal bandwidth is also slightly smaller. When the prediction horizon is larger than 1, although the QPI with under-smoothing bandwidth has a slightly larger LEN, its CVR is notably better than QPI with optimal bandwidth. Here, we do more simulation studies to show that the QPIs with optimal bandwidth and under-smoothing bandwidth are asymptotically equivalent. We do simulations with \cref{model1} and take $T+1$ to be 1000. The CVR and LEN of different QPIs are tabulated in \cref{Tab:model1T1000}.  

\begin{table}[htbp]
\centering
  \caption{The CVR and LEN of QPIs with 1000 sample on \cref{model1}}
  \vspace{2pt}
  \label{Tab:model1T1000}
\begin{tabular}{lcccccccccc}
  \toprule 
 Model 1: & \multicolumn{10}{c}{$X_t = \log(X_{t-1}^2 + 1) + \epsilon_t, \epsilon_t\sim N(0,1)$} \\
 \midrule
  & \multicolumn{5}{c}{CVR for each step} & \multicolumn{5}{c}{LEN for each step}\\
    $T = 1000$  & 1     & 2     & 3     & 4     & 5 & 1 & 2 & 3 & 4 & 5  \\[3pt]
QPI-f & 0.950 & 0.940 & 0.948 & 0.947 & 0.939 &3.86 & 4.50 & 4.66 & 4.70 & 4.71 \\QPI-f-u & 0.947 & 0.943 & 0.952 & 0.954 & 0.946& 3.86 & 4.56 & 4.74 & 4.79 & 4.81 \\ 
QPI-p & 0.949 & 0.938 & 0.951 & 0.951 & 0.943& 3.91 & 4.54 & 4.71 & 4.75 & 4.76 \\
QPI-p-u & 0.951 & 0.947 & 0.954 & 0.956 & 0.950 & 3.90 & 4.62 & 4.80 & 4.84 & 4.86 \\  
       \bottomrule
    \end{tabular}\\
    %   \tiny
    %   \raggedright
    %  \textit{Note:} 
\end{table}
Although the LEN of QPI with optimal bandwidth is always less than the variant with the under-smoothing bandwidth, the difference is marginal. In addition, these two types of QPIs have indistinguishable performance according to the CVR, which implies the asymptotic equivalence of applying optimal bandwidth or under-smoothing bandwidth. It also implies that adopting fitted or predictive residuals is also asymptotically equivalent.  
\clearpage

\section*{\textsc{Appendix C: The effects of applying under-smoothing or over-smoothing bandwidth on PPI}}\label{Appendix:diffoverunderband}

To see the effects of applying under-smoothing or over-smoothing tricks on the performance of PPI. We take sample size $T+1$ to be $50$ or $500$ and perform simulations 5000 times on the first model. Simulation results are shown below:

\begin{table}[htbp]
\centering
  \caption{The CVR and LEN of PPIs with under-smoothing or over-smoothing bandwidth strategies on \cref{model1}}
  \vspace{2pt}
  \label{Tab:model1PPIcompare}
\begin{tabular}{lcccccccccc}
  \toprule 
 Model 1: & \multicolumn{10}{c}{$X_t = \log(X_{t-1}^2 + 1) + \epsilon_t, \epsilon_t\sim N(0,1)$} \\
 \midrule
  & \multicolumn{5}{c}{CVR for each step} & \multicolumn{5}{c}{LEN for each step}\\
    $T = 500$ & 1     & 2     & 3     & 4     & 5 & 1 & 2 & 3 & 4 & 5  \\[3pt]
$L_2$-PPI-f-u & 0.943 & 0.940 & 0.945 & 0.943 & 0.948 & 3.88 & 4.54 & 4.71 & 4.77 & 4.78 \\ 
 $L_1$-PPI-f-u & 0.942 & 0.941 & 0.946 & 0.947 & 0.949 & 3.89 & 4.55 & 4.72 & 4.78 & 4.80\\ 
  $L_2$-PPI-p-u  & 0.946 & 0.949 & 0.947 & 0.952 & 0.954& 3.96 & 4.63 & 4.79 & 4.85 & 4.8  \\ 
 $L_1$-PPI-p-u & 0.946 & 0.950 & 0.947 & 0.951 & 0.954 & 3.97 & 4.64 & 4.81 & 4.86 & 4.88  \\ 
  $L_2$-PPI-f-o & 0.942 & 0.926 & 0.916 & 0.915 & 0.923 & 3.86 & 4.26 & 4.33 & 4.34 & 4.35 \\ 
  $L_1$-PPI-f-o & 0.943 & 0.925 & 0.921 & 0.918 & 0.922 & 3.87 & 4.27 & 4.34 & 4.36 & 4.36 \\ 
  $L_2$-PPI-p-o & 0.948 & 0.929 & 0.927 & 0.927 & 0.925 & 3.94 & 4.34 & 4.42 & 4.43 & 4.43 \\ 
  $L_1$-PPI-p-o & 0.949 & 0.931 & 0.928 & 0.925 & 0.924 & 3.95 & 4.35 & 4.43 & 4.44 & 4.44\\ 
    SPI & 0.946 & 0.947 & 0.948 & 0.950 & 0.956 & 3.89 & 4.57 & 4.76 & 4.82 & 4.84  \\   \\[3pt]
    $T = 50 $       &       &       &       &       &  \\[3pt]
  $L_2$-PPI-f-u & 0.912 & 0.919 & 0.919 & 0.925 & 0.931 & 3.95 & 4.53 & 4.67 & 4.72 & 4.74 \\ 
  $L_1$-PPI-f-u & 0.913 & 0.921 & 0.919 & 0.928 & 0.931 & 3.96 & 4.55 & 4.69 & 4.74 & 4.76 \\ 
  $L_2$-PPI-p-u & 0.943 & 0.945 & 0.942 & 0.946 & 0.950 & 4.38 & 4.95 & 5.08 & 5.12 & 5.14  \\ 
  $L_1$-PPI-p-u & 0.944 & 0.946 & 0.943 & 0.948 & 0.950 & 4.39 & 4.98 & 5.10 & 5.15 & 5.16  \\ 
  $L_2$-PPI-f-o & 0.911 & 0.880 & 0.869 & 0.869 & 0.873   & 3.78 & 3.93 & 3.96 & 3.97 & 3.97 \\ 
  $L_1$-PPI-f-o & 0.912 & 0.882 & 0.868 & 0.868 & 0.871  & 3.79 & 3.95 & 3.98 & 3.98 & 3.98  \\ 
  $L_2$-PPI-p-o & 0.940 & 0.918 & 0.903 & 0.908 & 0.910  & 4.20 & 4.37 & 4.40 & 4.41 & 4.42  \\ 
  $L_1$-PPI-p-o & 0.941 & 0.919 & 0.902 & 0.909 & 0.909 & 4.22 & 4.39 & 4.42 & 4.43 & 4.43 \\ 
  SPI & 0.950 & 0.947 & 0.946 & 0.947 & 0.950  & 3.89 & 4.58 & 4.76 & 4.82 & 4.84  \\ 
       \bottomrule
    \end{tabular}\\
      \raggedright
     \textit{Note:}  ``-o'' indicates the corresponding PPI is built with over-smoothing bandwidth on generating bootstrap series. 
\end{table}
The above results coincide with \cref{Corollary:1steppertinent}, i.e., both bandwidth strategies can give one-step ahead PPIs with satisfied CVR even for a small sample size with predictive residuals. The implication of \cref{Theorem:PPI_part2} is also verified, i.e., taking the under-smoothing bandwidth can keep the CVR at a high level for multi-step ahead predictions when the sample size is small. In addition, as the sample size increases, the CVR of PPI with over-smoothing bandwidth increases also. This phenomenon is guaranteed by the asymptotically valid property of PPI with no matter over-smoothing or under-smoothing bandwidth; see \cref{Theorem:PPI_part1}. 

\newpage
\section*{\textsc{Appendix D: The comparison of applying under-smoothing or optimal bandwidth on estimating the variance function for building PPI}}\label{Appendix:optbandwidthonestvar}

From \hyperref[Appendix:diffoverunderband]{Appendix C}, we have seen the advantage of applying under-smoothing bandwidth to estimate the model in the real and bootstrap worlds when the model is with homoscedastic errors. For the model with heteroscedastic errors, as we have mentioned in \cref{Subsec:PPIHeter}, we can rely on the optimal bandwidth to estimate the variance functions. 

To check this claim, we consider two strategies on the bandwidth of the estimator for the variance function: (1) Take the under-smoothing bandwidth as we do for the mean function estimator; (2) Take the bandwidth with optimal rate. For estimating the mean function in the bootstrap world, we keep using the under-smoothing bandwidth strategy. Simulation results based on \cref{model2} with a small sample size are shown below:
\begin{table}[htbp]
\centering
  \caption{The CVR and LEN of PPIs with two strategies on estimating the variance function}
  \vspace{2pt}
  \label{Tab:model2PPIcompare_opvariance}
\begin{tabular}{lcccccccccc}
  \toprule 
 Model 1: & \multicolumn{10}{c}{$X_t = \sin(X_{t-1}) + \epsilon_t\sqrt{0.5 + 0.25X_{t-1}^2}, \epsilon_t\sim N(0,1)$} \\
 \midrule
  & \multicolumn{5}{c}{CVR for each step} & \multicolumn{5}{c}{LEN for each step}\\
    $T = 50, \text{Rep} = 5000$        &       &       &       &       &  \\[3pt]
  $L_2$-PPI-f-u & 0.871 & 0.896 & 0.921 & 0.915 & 0.923 & 3.50 & 4.24 & 4.41 & 4.48 & 4.52  \\ 
  $L_1$-PPI-f-u & 0.877 & 0.901 & 0.919 & 0.918 & 0.925 & 3.52 & 4.24 & 4.42 & 4.49 & 4.53\\ 
  $L_2$-PPI-p-u & 0.925 & 0.939 & 0.946 & 0.946 & 0.946 & 4.82 & 5.47 & 5.63 & 5.71 & 5.81  \\ 
  $L_1$-PPI-p-u & 0.927 & 0.935 & 0.945 & 0.949 & 0.949& 4.80 & 5.39 & 5.51 & 5.65 & 5.75  \\ 
  $L_2$-PPI-f-opv & 0.885 & 0.891 & 0.923 & 0.920 & 0.918 & 3.45 & 4.12 & 4.34 & 4.39 & 4.43\\ 
  $L_1$-PPI-f-opv  & 0.885 & 0.893 & 0.927 & 0.919 & 0.917 & 3.47 & 4.14 & 4.36 & 4.41 & 4.45  \\ 
  $L_2$-PPI-p-opv & 0.934 & 0.939 & 0.947 & 0.950 & 0.947& 4.75 & 5.28 & 5.49 & 5.56 & 5.60   \\ 
  $L_1$-PPI-p-opv& 0.940 & 0.940 & 0.946 & 0.951 & 0.943 & 4.72 & 5.21 & 5.40 & 5.45 & 5.55 \\ 
  SPI & 0.943 & 0.939 & 0.958 & 0.945 & 0.945 & 3.38 & 4.11 & 4.33 & 4.38 & 4.40 \\ 
       \bottomrule
    \end{tabular}\\
      \raggedright
     \textit{Note:}  ``-opv'' indicates the corresponding PPI is built by  optimal bandwidth for the variance function estimator. 
\end{table}

From \cref{Tab:model2PPIcompare_opvariance}, the LEN of PPI with optimal bandwidth on estimating variance function is always smaller than the corresponding PPI with under-smoothing bandwidth. At the same time, the CVR of both types of PPI is indistinguishable for $k>1$. For the one-step ahead prediction, the former PPI is notably better than the latter PPI. This phenomenon is implied by \cref{Remark:underopbandwidthdiss}, i.e., the best strategy for one-step ahead PPI is choosing bandwidths with the optimal rate for both estimators of mean and variance functions.

\end{document}